\documentclass[structabstract]{aa}  
\pdfoutput=1  %AÑADIR PARA ASTROPH!!!

\usepackage{amsmath} 
\usepackage{graphicx}
\usepackage[]{natbib}
\DeclareGraphicsExtensions{.jpg,.mps,.pdf,.png,.eps,.ps} % Formats d’images
\usepackage{txfonts}
\usepackage{url}
\usepackage{longtable}
\begin{document}

\title{TNOs are Cool: a survey of the Transneptunian Region XII}

\subtitle{Thermal light curves of Haumea, 2003 VS$_{2}$ and 2003 AZ$_{84}$ with Herschel Space Observatory-PACS\thanks{Herschel is an ESA space observatory with science instruments provided by European--led Principal Investigator consortia and with important participation from NASA. PACS: The Photodetector Array Camera and Spectrometer is one of Herschel's instruments.}}

   \author{P. Santos-Sanz\inst{1}
                   \and 
                   E. Lellouch\inst{2}
                   \and
                   O. Groussin\inst{3}
                   \and            
                   P. Lacerda\inst{4}
                   \and
                   T.G. M\"uller\inst{5}
                   \and            
                   J.L. Ortiz\inst{1}
                   \and            
                   C. Kiss\inst{6}   
                   \and            
                   E. Vilenius\inst{5,7}
                   \and            
                   J. Stansberry\inst{8}
                   \and
                   R. Duffard\inst{1}
                   \and                    
                   S. Fornasier\inst{2,9}
                   \and            
                   L. Jorda\inst{3}
                   \and
                   A. Thirouin\inst{10}
                   }

\offprints{P. Santos-Sanz: psantos@iaa.es}

\institute{
Instituto de Astrof\'{\i}sica de Andaluc\'{\i}a (CSIC), Glorieta de la Astronom\'{\i}a s/n, 18008-Granada, Spain.
\email{psantos@iaa.es} 
\and
LESIA-Observatoire de Paris, CNRS, UPMC Univ. Paris 6, Univ. Paris-Diderot, France.
\and
Aix Marseille Universit\'e, CNRS, LAM (Laboratoire d'Astrophysique de Marseille) UMR 7326, 13388, Marseille, France.
\and
Astrophysics Research Centre, Queen's University Belfast, Belfast BT7 1NN, United Kingdom.
\and
Max--Planck--Institut f\"ur extraterrestrische Physik (MPE), Garching, Germany.
\and
Konkoly Observatory of the Hungarian Academy of Sciences, Budapest, Hungary.
\and 
 Max--Planck-Institut f\"ur Sonnensystemforschung (MPS), Justus-von-Liebig-Weg 3, D-37077 G\"ottingen, Germany.
\and
 Space Telescope Science Institute, 3700 San Martin Drive, Baltimore, MD 21218, USA.
\and
 Univ. Paris Diderot, Sorbonne Paris Cit\'{e}, 4 rue Elsa Morante, 75205 Paris, France.
\and
 Lowell Observatory, 1400 W Mars Hill Rd, Flagstaff, Arizona, 86001, USA.
}
   
   \date{Received  / Accepted }

% \abstract{}{}{}{}{} 
% 5 {} token are mandatory
 
  \abstract
  % context heading (optional)
  % {} leave it empty if necessary  
   {Time series observations of the dwarf planet Haumea and the Plutinos 2003 VS$_{2}$ and 2003 AZ$_{84}$ with Herschel/PACS are presented in this work. Thermal emission of these trans-Neptunian objects (TNOs) were acquired as part of the TNOs are Cool Herschel Space Observatory key programme. }
  % aims heading (mandatory)
   {We search for the thermal light curves at 100 and 160 $\mu$m of Haumea and 2003 AZ$_{84}$, and at 70 and 160 $\mu$m for 2003 VS$_{2}$ by means of photometric analysis of the PACS data. The goal of this work is to use these thermal light curves to obtain physical and thermophysical properties of these icy Solar System bodies. 
   }
  % methods heading (mandatory)
   {When a thermal light curve is detected, it is possible to derive or constrain the object thermal inertia, phase integral and/or surface roughness with thermophysical modeling.
   }
  % results heading (mandatory)
   {Haumea's thermal light curve is clearly detected at 100 and 160 $\mu$m. The effect of the reported dark spot is apparent at 100 $\mu$m. Different thermophysical models were applied to these light curves, varying the thermophysical properties
of the surface within and outside the spot. Although no model gives a perfect fit to the thermal observations, results imply  an extremely low thermal inertia ($<$ 0.5 J m$^{-2}$ s$^{-1/2}$ K$^{-1}$, hereafter MKS) and a high phase integral ($>0.73$) for Haumea's surface. We note that the dark spot region appears to be only weakly different from the rest of the object, with modest changes in thermal inertia and/or phase integral. The thermal light curve of 2003 VS$_{2}$ is not firmly detected at 70 $\mu$m and at 160 $\mu$m but a thermal inertia of (2$\pm$0.5) MKS can be derived from these data. The thermal light curve of 2003 AZ$_{84}$ is not firmly detected at 100 $\mu$m. We apply a thermophysical model to the mean thermal fluxes and to all the Herschel/PACS and Spitzer/MIPS thermal data of 2003 AZ$_{84}$, obtaining a close to pole-on orientation as the most likely for this TNO.
}
  % conclusions heading (optional), leave it empty if necessary 
   {For the three TNOs, the thermal inertias derived from light curve analyses or from the thermophysical analysis of the mean thermal fluxes confirm the generally small or very small surface thermal inertias of the TNO population, which is consistent with a statistical mean value $\Gamma _{mean} =$ 2.5 $\pm$ 0.5 MKS.
}
   
\titlerunning{PACS thermal light curves of Haumea, 2003 AZ$_{84}$ and 2003 VS$_{2}$}

   \maketitle

%
%________________________________________________________________

\section{Introduction}

The study of the visible photometric variation of  Solar System minor bodies enables us to determine optical light curves (flux or magnitude versus time), for which essential parameters are the peak-to-peak amplitude and the rotational period of the object. Short-term photometric variability of TNOs and Centaurs can be shape-driven \citep[i.e. Jacobi-shaped rotating body, such as Varuna, ][]{2002AJ....123.2110J} or be causes by albedo contrasts on the surface of a Maclaurin-shaped rotating spheroid (e.g. the Pluto case). Combinations of shape and albedo effects are also possible and very likely occur within TNOs (e.g. Makemake) and Centaurs. Contact-binaries can also produce short-term photometric variability within the TNO and Centaur populations. The largest amplitudes are usually associated with Jacobi shapes and the smallest ones with Maclaurin shapes with highly variegated surfaces. Large amplitudes can also be associated with contact-binaries and small amplitudes with objects with rotational axes close to pole-on. If we know the rotational properties (i.e. rotation period and amplitude) of a Jacobi shaped object, it is possible to derive the axes ratio of the ellipsoid (i.e. a shape model) and also a lower limit for the density \citep[][ and references therein]{2002AJ....123.2110J,2008ssbn.book..129S}, assuming the object is in hydrostatic equilibrium \citep{Chandra87} with a certain aspect angle. On the other hand, if we suspect that the object has a Maclaurin shape we can derive a shape model from the rotational period, but it is needed to assume a realistic density in this case. The majority of the TNOs/Centaurs ($\sim$ 70\%) present shallow light curves (amplitudes less than 0.15 magnitudes), which indicates that most of them are Maclaurin-shaped bodies \citep{2009A&A...505.1283D,2010A&A...522A..93T}. For a couple of special cases (only for Centaurs until now) where more information is available (i.e. long-term changes in light curve amplitudes), the position of the rotational axis can be derived or at least constrained \citep{2005Icar..175..390T,2014A&A...568A..79D,Ortiz2015,estela2016}.

Complementary to optical light curves, thermal light curves are a powerful tool to obtain additional information about physical and, in particular, thermal properties of these bodies. At first order, immediate comparison of the thermal and optical light curves enables us to differentiate between shape-driven light curves (the thermal light curve is then correlated with the optical one) and light curves that are the result of  albedo markings (the two light curves are anti-correlated). Furthermore, quantitative modeling of the thermal light curve enables us to constrain the surface energetic and thermal properties, namely its bolometric albedo, thermal inertia, and surface roughness \citep[e.g. Pluto, see][ and references therein]{LellouchPlutoLCs}. In a more intuitive way, in the case of positively correlated thermal and optical light curves, it is also possible to constrain the thermal inertia using the delay between the thermal and optical light curves and their relative
amplitudes.

Until  recently, only a few thermal light curves of outer Solar System minor bodies (or dwarf planets) have been obtained. Pluto's thermal light curve was detected with ISO/PHOT, Spitzer/MIPS, and /IRC at a variety of wavelengths longwards of 20 $\mu$m \citep{2000Icar..147..220L,2011Icar..214..701L}. More recently, Pluto was also observed with the Herschel Space Observatory \citep{2010A&A...518L...1P} Photodetector Array Camera and Spectrometer \citep[PACS:][]{2010A&A...518L...2P} and with the Spectral and Photometric Imaging
Receiver \citep[SPIRE:][]{2010A&A...518L...3G}, which provides thermal light curves at six wavelengths: 70, 100, 160, 250, 350, and 500 $\mu$m \citep{LellouchPlutoLCs}. 
A marginal thermal light curve of dwarf planet Haumea was reported with Spitzer-MIPS \citep{2009DPS....41.6506L,2014EM&P..tmp....3L}. The definite
detection of Haumea's thermal light curve was obtained with Herschel/PACS within the Science Demonstration Phase \citep[SDP,][]{2010A&A...518L.147L}. Other tentative thermal light curves of TNOs/Centaurs that were observed with Herschel/PACS (i.e. Varuna, Quaoar, Chiron, Eris) were also   presented for the first time \citep{SantosSanz2014} and will be published separately \citep[e.g.][]{kiss2016}.

Here, we present thermal time series photometry of the dwarf planet Haumea and the Plutinos 2003 VS$_{2}$ and 2003 AZ$_{84}$ that were taken with Herschel/PACS using its 3-filter bands, which are centred at 70, 100, and 160 $\mu$m (hereafter blue, green, and red bands, respectively). In the case of Haumea, we present additional and improved data and we merge them with the SDP observations that were originally presented in \cite{2010A&A...518L.147L}. The thermal time series of 2003 VS$_{2}$ and 2003 AZ$_{84}$ are presented here for the first time. % and thermal light curves of these objects never have been observed until now.
 The Herschel/PACS observations are described in Sect. \ref{observations}, the data reduction and photometry techniques applied are detailed in Sect. \ref{reductionAndPhotom}. Data are analyzed, modeled, and interpreted in Sect. \ref{Discuss}.
Finally, the major conclusions of this work are summarized and discussed in Sect. \ref{summary}.

%__________________________________________________________________

\section{Observations}
\label{observations}

The observations presented here are part of the project TNOs are Cool: a survey of the trans-Neptunian region, a Herschel Space Observatory open time key programme \citep{2009EM&P..105..209M}. This programme  used $\sim$ 372 hours of Herschel time (plus $\sim$ 30 hours within the SDP) to observe 130 TNOs/Centaurs, plus two giant planet satellites (Phoebe and Sycorax), with Herschel/PACS; 11 of these objects were also observed with Herschel/SPIRE at 250, 350, and 500 $\mu$m \citep[see][]{2013A&A...555A..15F}, with the main goal of obtaining sizes, albedos, and thermophysical properties of a large set of objects that are representative of the different dynamical populations within the TNOs. For PACS measurements, we used a range of observation durations from about 40 to 230 minutes based on flux estimates for each object. Results of the PACS and SPIRE measurements to date have been published in \cite{2010A&A...518L.146M}, \cite{2010A&A...518L.147L}, \cite{2010A&A...518L.148L}, \cite{2012A&A...541A..92S}, \cite{2012A&A...541A..93M}, \cite{2012A&A...541A..94V}, \cite{2012A&A...541L...6P}, \cite{2013A&A...555A..15F}, \cite{2013A&A...557A..60L}, \cite{2014A&A...564A..35V}, \cite{2014A&A...564A..92D}\footnote{All the TNOs are Cool results (and additional information) are collected in the public web page:  \url{http://public-tnosarecool.lesia.obspm.fr}}. In addition,
four bright objects (Haumea, 2003 VS$_{2}$, 2003 AZ$_{84}$ and Varuna) were re-observed long enough to search for thermal emission variability related to rotation (i.e. thermal light curve). This paper presents 
results for the first three. Other objects (Pluto, Eris, Quaoar, Chiron)  were also observed  outside of the TNOs are Cool programme to search for their thermal light curve. 

The general strategy we used to detect a thermal light curve with Herschel/PACS was to perform a long observation covering most of the expected light curve duration, followed by a shorter follow-on observation, which enabled us to  clean the images' backgrounds, as explained in Sect \ref{Datared}. 

Dwarf planet (136108) Haumea was observed twice with Herschel/PACS in mini scan maps mode at 100 and 160 $\mu$m. The first visit was performed on 23 December 2009, which covers 86\% of its 3.92 h rotational period followed by a shorter follow-on observation on 25 December 2009. The second one was obtained on 20 June 2010 (follow-on observations on 21 June 2010) using the same detector and bands and covering 110\% of its rotational period.

The Plutino (84922) 2003 VS$_{2}$ was observed with Herschel/PACS in mini scan-maps mode at 70 and 160 $\mu$m on 10 August 2010, covering 106\% of its 7.42 hr rotational period. Follow-on observations were performed on 11 August 2010.

The binary Plutino (208996) 2003 AZ$_{84}$ was observed with Herschel/PACS in mini scan maps mode at 100 and 160 $\mu$m on 26-27 September 2010 (with follow-on observations on 28 September 2010). The observation lasted $\sim$7.4 h, i.e 110\% of an assumed single-peak rotational period of 6.79 h (or 55 \%
of a double-peaked period of 13.58 h).

All observations were made using only one scanning direction %of the scan angles: 70$\degr$ or 110$\degr$ 
(see e.g. \citeauthor{2012A&A...541A..92S} \citeyear{2012A&A...541A..92S} for a detailed description of the mini scan maps mode in the case of TNOs\footnote{The observing mode itself is described in the technical note PACS Photometer-Point/Compact Source Observations: Mini Scan-Maps $\&$ Chop-Nod, 2010, PICC-ME-TN-036, custodian T. Muller}).

Table \ref{tableOBS} shows the orbital parameters, B-R colors, absolute magnitudes, rotational properties, taxonomy, and dynamical classification of the observed objects. This table also includes the radiometric
solutions of these three objects (equivalent diameter for an equal-area sphere, geometric albedo, beaming factor) previously published as part of the TNOs are Cool project. Table \ref{tableOBSIDs} shows the observational circumstances of each one of these TNOs. 

Owing to the spatial resolution of Herschel/PACS, the satellites of the binaries/multiple systems (i.e. Haumea and 2003 AZ$_{84}$) are not resolved, and their thermal fluxes are merged with the thermal flux of the main body.

\scriptsize
\begin{table*}

\caption{Orbital parameters, absolute magnitudes, B-R colors, photometric variation, taxonomy, dynamical classification, and previously published Herschel results of the observed objects} % title of Table

\label{tableOBS} % is used to refer this table in the text

\centering % used for centering table
\scalebox{0.77}{%
\begin{tabular}{l c c c c l c c c c c c c c} % centered columns (4 columns)

\hline\hline % inserts double horizontal lines

Object  &       a       &       q       &       i       &       e       &       H$_{V}$ &       B-R & P     &       $\Delta$m       &       Taxon.  &       Class. & D & p$_{V}$ & $\eta$ \\
        &       [AU]    &       [AU]    &       [deg]   &               &[mag]& [mag]&  [h]     &       [mag]   &               &       & [km] & [\%] & \\

% table heading

\hline % inserts single horizontal line

%(15874) 1996 TL$_{66}$
(136108) Haumea$^{*}$   &       43.34 & 35.14   &       28.2    &       0.19    &       0.428$\pm$0.011$^{a}$ & 1.00$\pm$0.03$^{b,e,f}$       &       3.915341$\pm$0.000005$^{g}$ &   0.28$\pm$0.02$^{a}$         & BB &  Res     &       1240$^{+69}_{-58}$ $^{i}$ & 80.4$^{+6.2}_{-9.5}$  $^{i}$ & 0.95$^{+0.33}_{-0.26}$  $^{i}$ \\
(84922) 2003 VS$_{2}$   &       39.38   &       36.45   &       14.8    &       0.07    &       4.110$\pm$0.380$^{c}$ & 1.52$\pm$0.03$^{c}$
        &       7.4175285$\pm$ 0.00001$^{h}$    &       0.21$\pm$0.01$^{a}$ &       BB      &       Plu     &       523$^{+35}_{-34}$ $^{j}$ & 14.7$^{+6.3}_{-4.3}$ $^{j}$ & 1.57$^{+0.30}_{-0.23}$ $^{j}$ \\
(208996) 2003 AZ$_{84}^{*}$     &       39.60   &       32.55   &       13.6         &       0.18    &       3.760$\pm$0.058$^{c, d}$ & 1.05$\pm$0.06$^{c}$
&       6.7874$\pm$0.0002$^{h}$ &       0.07$\pm$0.01$^{a}$     &       BB         &       Plu     &       727$^{+62}_{-67}$ $^{j}$ & 10.7$^{+2.3}_{-1.6}$ $^{j}$ & 1.05$^{+0.19}_{-0.15}$ $^{j}$ \\

\hline %inserts single line

\end{tabular}}

\begin{flushleft}

\footnotesize{ \textbf{*} Indicates that the object is a known binary/multiple system. \textbf{Orbital parameters:} \textbf{(a)} semimajor axis in Astronomical Units (AU), \textbf{(q)} perihelion distance in AU, \textbf{(i)} orbital inclination in degrees, and \textbf{(e)} eccentricity, from Minor Planet Center (MPC-IAU) database, July 2016. \textbf{H$_{V}$ [mag]:} average visual magnitude obtained from papers referenced below. \textbf{B-R [mag]} colors. \textbf{P [h]} preferred single or double-peaked rotational period. \textbf{$\Delta$m [mag]:} light curve peak-to-peak amplitude. \textbf{Taxon.}: taxonomic color class \citep[][and references therein]{2010A&A...510A..53P}. \textbf{Class.}: dynamical classification following \cite{2008ssbn.book...43G} scheme: Res (Resonant), Plu (Plutino). \textbf{Herschel results:} \textbf{(D)} area-equivalent diameter, \textbf{(p$_{V}$)} geometric albedo at V-band, \textbf{($\eta$)} beaming factor determined from NEATM thermal modeling. \textbf{References:} $^{a)}$ \cite{2010A&A...522A..93T}; $^{b)}$ \cite{2007AJ....133...26R}; $^{c)}$ \cite{2012A&A...541A..93M} and references therein; $^{d)}$ \cite{2010A&A...510A..53P}; $^{e)}$ \cite{2007AJ....134.2046J}; $^{f)}$ \cite{2007ApJ...655.1172T}; $^{g)}$ \cite{2010A&A...518L.147L}; $^{h)}$ This work (a further description of the observations and techniques leading to these rotation periods are detailed in \citeauthor{thirouin2013} \citeyear{2012A&A...541A..92S}); $^{i)}$ \cite{2013A&A...555A..15F}; $^{j)}$ \cite{2012A&A...541A..93M}
}

\end{flushleft}

\end{table*}

\normalsize

\begin{table*}

\caption{Individual observational circumstances} % title of Table

\label{tableOBSIDs} % is used to refer this table in the text
 
\centering % used for centering table

\begin{tabular}{r c c c c c c c c} % centered columns (4 columns)

\hline\hline % inserts double horizontal lines

Object  &       OBSIDs  & Band  & Dur. [min]    & Covered [\%] &        Mid-time        &       $r_h$ [AU]    &       $\Delta$[AU]    &       $\alpha$[deg]   \\

% table heading

\hline % inserts single horizontal line

(136108) Haumea              &  1342188470$^{*}$     & g/r      &       201.4 &   86 &        23-Dec-2009 07:32:43  & 51.0279 & 51.2615 &     1.08\\
                                                                                &       1342188520$^{\dagger}$     & g/r       &        40.3   &      &        25-Dec-2009 06:33:48  & 51.0276 & 51.2317 &     1.09\\
                                                                                &       1342198851                    & g/r      & 258.6 &  110   &      20-Jun-2010     22:54:28  &     51.0012 & 50.7370 &     1.12\\
                                                                                &       1342198903-04$^{\dagger}$  & b/r  &  20.0 &      &        21-Jun-2010 22:52:00  & 51.0010 & 50.7514 &       1.12\\
                                                                                &       1342198905-06$^{\dagger}$  & g/r  &  20.0 &      &        21-Jun-2010 23:13:02  & 51.0010 & 50.7516 &       1.12\\                                          
(84922) 2003 VS$_{2}$                   &       1342202371               & b/r   & 470.1 &  106 &        10-Aug-2010 13:30:28      &     36.4761 & 36.8208 &       1.50\\
                                                                                &       1342202574-75$^{\dagger}$          & b/r  & 20.0 &     &  11-Aug-2010 03:12:38      &     36.4761 & 36.8119 &       1.51\\
                                                                                &       1342202576-77$^{\dagger}$          & g/r  & 20.0 &     &  11-Aug-2010 03:33:40      &     36.4761 & 36.8116 &       1.51\\
(208996) 2003 AZ$_{84}$                 &       1342205152               & g/r   & 446.6 &  110   &      27-Sep-2010 03:36:40      &     45.3011 & 45.6719 &       1.18\\
                                                                                &       1342205222-23$^{\dagger}$        & b/r   & 20.0 &         &      28-Sep-2010 03:01:13      &     45.3008 & 45.6561 &       1.19\\
                                                                                &       1342205224-25$^{\dagger}$        & g/r   & 20.0 &         &      28-Sep-2010 03:22:15      &     45.3008 & 45.6559 &       1.19\\
\hline %inserts single line

\end{tabular}

\footnotesize{\textbf{OBSIDs:} \textit{Herschel} internal observation IDs. $^{*}$Observations made during Herschel Science Demonstration Phase (SDP). $^{\dagger}$Follow-on observations used to apply the background subtraction techniques. \textbf{Band:} PACS filter used during OBSID, b stands for blue (70 $\mu$m), g stands for green (100 $\mu$m) and r stands for red (160 $\mu$m). \textbf{Dur. [min]:} total duration of the observation in each band (70/160 $\mu$m or 100/160 $\mu$m). \textbf{Covered [\%]:} Percentage of the (preferred) rotational period covered by the observations (see Table \ref{tableOBS}). \textbf{Mid-time:} Mean date and UT time of the observation. \textbf{$r_h$ [AU]:} heliocentric distance at mid-time in AU. \textbf{$\Delta$ [AU]:} distance object-\textit{Herschel} at mid-time in AU. \textbf{$\alpha$[deg]:} phase angle in degrees at mid-time.}

\end{table*}

%______________________________________________________________

\section{Data reduction and photometry}
\label{reductionAndPhotom}

\subsection{Data reduction}
\label{Datared}

PACS images obtained from the Herschel Space Observatory were processed using the \textit{Herschel Interactive Processing Environment} (HIPE\footnote{HIPE is a joint development by the \textit{Herschel} Science Ground Segment Consortium, consisting of ESA, the NASA \textit{Herschel Science Center}, and the HIFI, PACS and SPIRE consortia members, see: http://herschel.esac.esa.int/DpHipeContributors.shtml}) and our own adapted pipelines developed within the TNOs are Cool project. The application of the pipeline provides individual or single maps, each one of these single images covering $\sim$4.7 minutes. The re-sampled pixel scale of the single maps is 1.1$\arcsec$/pixel, 1.4$\arcsec$/pixel, and 2.1$\arcsec$/pixel for the 70 $\mu m$ (blue), 100 $\mu m$ (green), and 160 $\mu m$ (red) bands, respectively. Apparent motion over the duration of
an individual map is negligible compared to the PACS PSF (FWHM in radius is 5.2$\arcsec$/7.7$\arcsec$/12$\arcsec$ in blue/green/red bands, respectively) and does not need to be corrected.  These single maps are combined afterwards, using ephemeris-based recentering processes within HIPE, to obtain enough signal-to-noise ratio (S/N) to perform a good photometry, while keeping enough time resolution to resolve the thermal light curve, in a similar way to \cite{2010A&A...518L.147L}. The exact number of individual and combined maps for each target is detailed in Sect. \ref{photom}.

\begin{figure}[!hpbt]
   \centering
   
   \includegraphics[width=9cm]{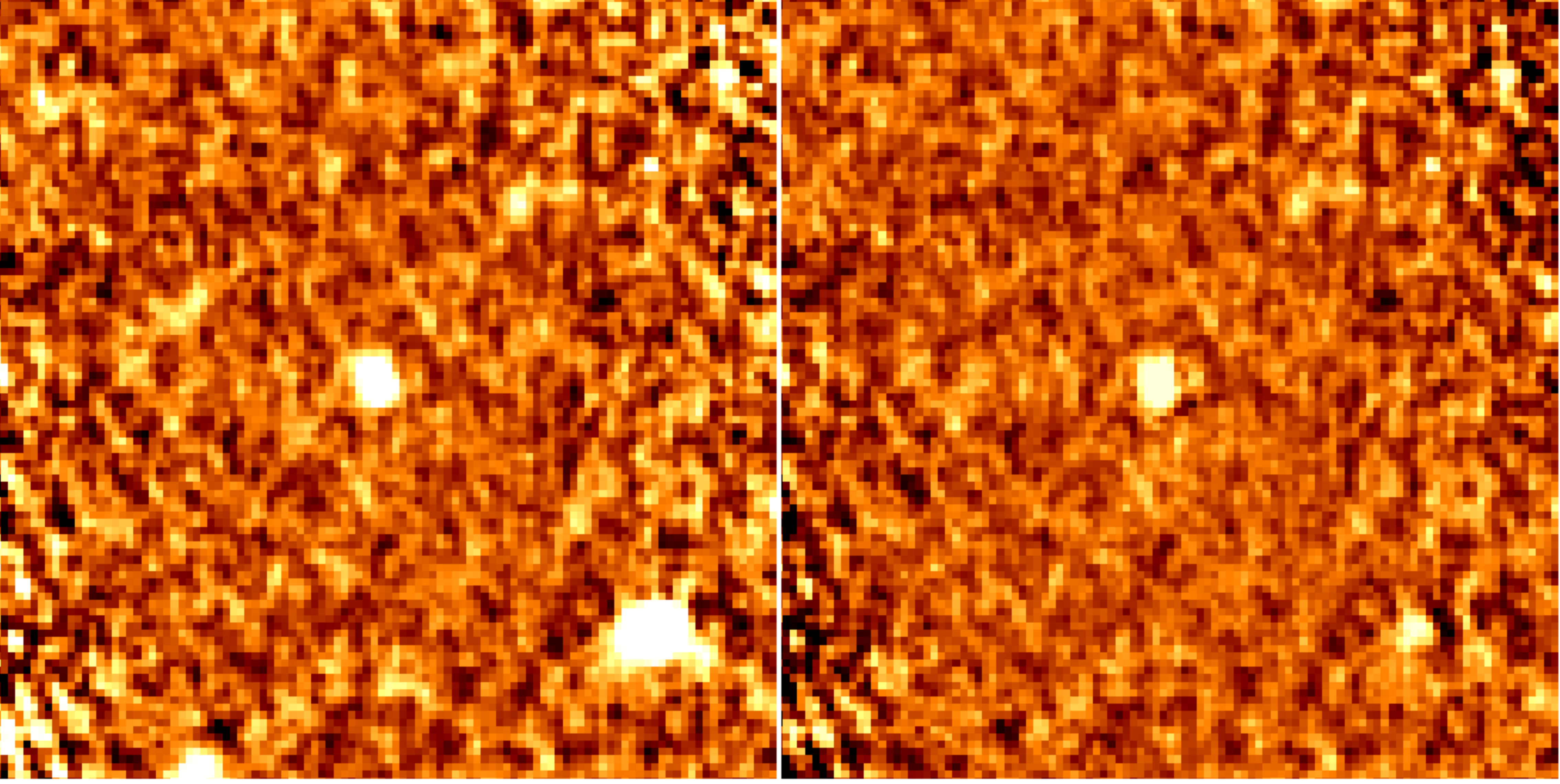}
   \includegraphics[width=9cm]{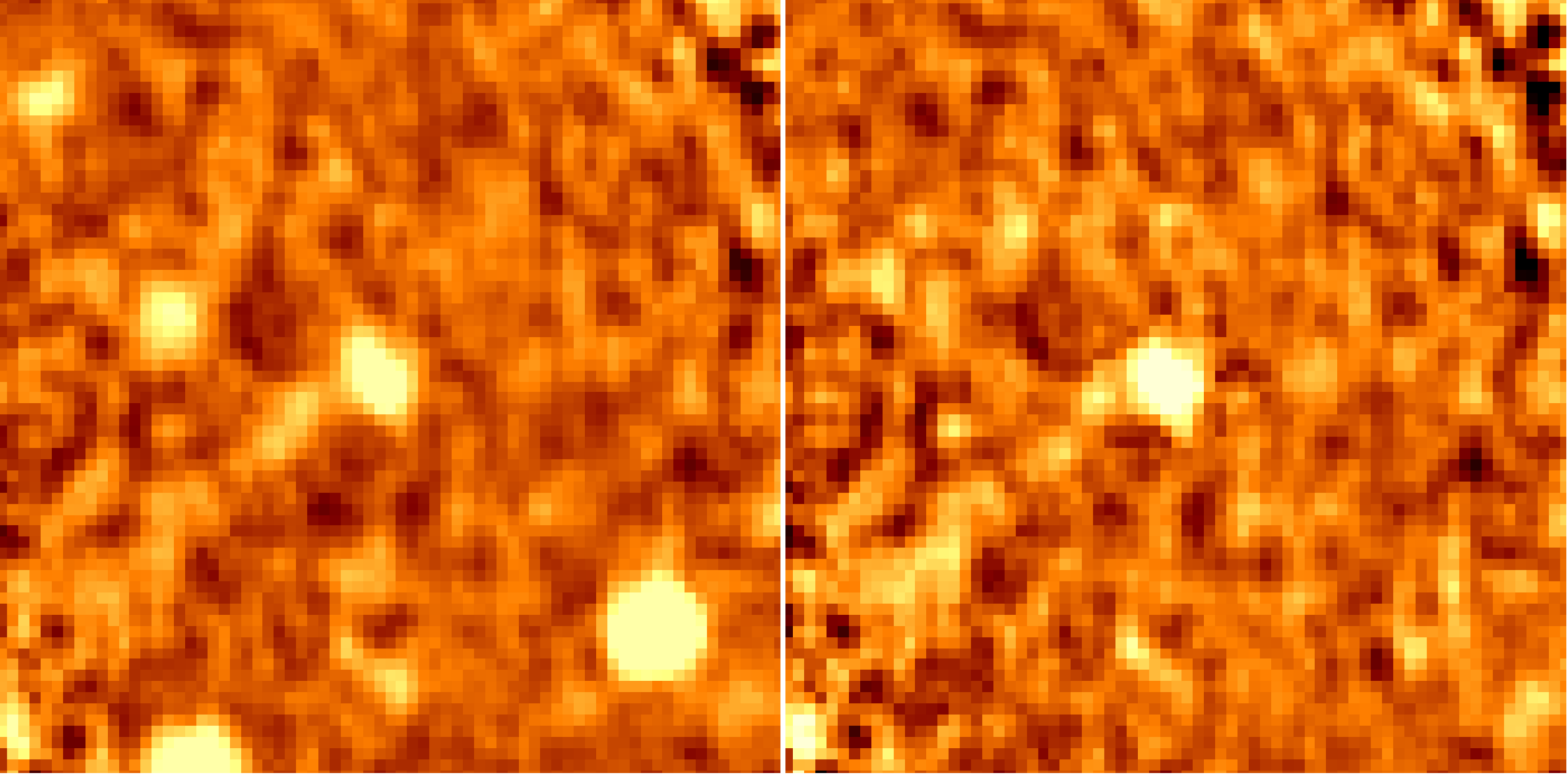}  
  
      \caption{Images (top line: 100 $\mu$m; bottom line: 160 $\mu$m) extracted from the 20 Jun 2010 light curve of Haumea. Left: original images. Right: background-subtracted images. The Field of View (FOV) is 2.5$\arcmin$ x 2.5$\arcmin$. Haumea is at centre. }
      
         \label{haumea_path}

\end{figure}

To minimize at best contamination by background sources, all light curve data are associated with complementary observations
acquired one or few days later (follow-on observations), where the target has moved enough that a background map can be determined
and subtracted from each combined map of the light curve. The method was
demonstrated in Spitzer/MIPS TNOs/Centaurs observations \citep{stansberry08,Brucker2009}. However, this technique to remove background sources fails when trying to remove some background features in the 2003 AZ$_{84}$ images. In this case another technique, known as double-differential background subtraction, is applied. A complete and detailed description of the data reduction process, the background subtraction and the double-differential techniques applied to the Herschel/PACS images can be found in \cite{2012A&A...541A..92S} and in \cite{2013ExA...tmp...36K}. Figures \ref{haumea_path}, \ref{2003VS2_path}, and \ref{2003AZ84_path} illustrate the advantages of these background-removing techniques for the three targets respectively.

%______________________________________________________________   
   
\subsection{Photometry}
\label{photom}

As indicated above, single maps obtained with a time resolution of 4.7 min were combined to improve image quality for photometry.
Typically, the number of single images to be combined was larger at 160 $\mu$m than at 70/100 $\mu$m, owing to lower S/N
and larger sky residuals. The details can be found below:

Haumea data were taken in two epochs, each time using the green (100 $\mu$m)/red (160 $\mu$m) filter combination. For the first
(resp. second) epoch, the total number of single images is 40 (resp. 55) per filter. These images were grouped
by 4 (18.8 minutes time resolution) in the green and by 6 (28.2 minutes per data point) in the red. 
The choice of this particular grouping of single images for the green (by 4) and the red (by 6) for the 
Haumea data is based on a compromise between obtaining enough S/N to extract a reliable photometry and having enough time 
resolution to properly sample the light curve, as mentioned above. Usually, more single images must be grouped for the red band 
because those images are normally noisier than images at other shorter wavelengths (even after the application of background-removing techniques).

The different grouping elections for 2003 VS$_{2}$ and 2003 AZ$_{84}$ are based on the same described compromise between S/N and time resolution. 
After removing clear outliers in the Haumea data, 37 images
remain for the first epoch (resp. 53 for the second epoch) at green band, and 35 
images for the first epoch (resp. 50 ) at red band (see Table \ref{HaumeaPACSdata} in Appendix \ref{append1}). 

2003 VS$_{2}$ was observed in blue/red combination, with 96 single maps for each color. We grouped these single images by 5 for the blue (23.5 minutes per data point) and by 10 for the red band (47.0 minutes per data point). Each blue data point is independent of the previous and following point (no time overlap) while, for the red, each data point has a time overlap of 23.5 minutes with the previous and following point. Consecutive points for both bands have a separation of $\sim$ 0.05 in rotational phase, clear data outliers have been removed. At the end, 18 data points remain for the blue band, and 18 for the red one (see Table \ref{VS2PACSdata} in Appendix \ref{append1}).

Similarly, 95 single maps of 2003 AZ$_{84}$ were acquired in green/red combination. They were grouped by 6 in the green without time overlap between previous and following point, and with a separation between consecutive points of $\sim$ 0.07 in phase. Final data points for the green are 15 (see Table \ref{AZ84PACSdata} in Appendix \ref{append1}). Red images were discarded for a thermal light curve analysis because they are very contaminated by background sources (even after the application of background removing techniques) and the final photometry on these images is very noisy. We still use the mean value of the red band flux (see Table \ref{thermal_data_az84}) for thermophysical model analysis.

Photometry was performed on the combined maps. The flux of the objects was obtained using DAOPHOT \citep{1987PASP...99..191S} routines adapted to IDL\footnote{Interactive Data Language, Research Systems Inc.}  to perform synthetic-aperture photometry on the final images. The object is usually located at the centre or very close to the centre of the images. Since our targets are bright enough, we do not need to use ephemeris coordinates to find them: photocentre routines are then used to obtain the best coordinates to place the centre of the circular aperture. Once the photocentre is obtained, we performed aperture photometry for radii that span from 1 to 15 pixels. We applied the aperture correction method \citep{1989PASP..101..616H} for each aperture radius using the tabulated encircled energy fraction for a point-source observed with PACS\footnote{M\"uller et al. (2011): PACS Photometer -Point-Source Flux Calibration, PICC-ME-TN-037, Version 1.0; Retrieved November 23, 2011; \url{http://herschel.esac.esa.int/twiki/pub/Public/PacsCalibrationWeb/pacs_bolo_fluxcal_report_v1.pdf}}. Uncertainties on the fluxes are estimated by means of a Monte-Carlo technique, in which artificial sources are randomly implanted on the images. We obtain and correct by aperture the fluxes of these artificial sources using a median optimum aperture radius. Uncertainties are computed as the standard deviation of these implanted fluxes. These photometric techniques and uncertainty estimations used to extract the PACS fluxes are further described in \cite{2012A&A...541A..92S} and \cite{2013ExA...tmp...36K}.

In addition to the random photometric errors, the data may suffer from a systematic flux calibration uncertainty, which is estimated
to be $\sim$5 \%. As the latter affects all points of a given light curve in an identical way, it is not included in the 
individual error bars. Color corrections, which are $\sim$1-2\% for Herschel/PACS data, were applied to the fluxes (see caption in Tables \ref{HaumeaPACSdata}, \ref{VS2PACSdata}, and \ref{AZ84PACSdata} for the exact value of the color-correction factors applied).

Finally, time-phasing of all the images was computed using the preferred rotational periods (see Table \ref{tableOBS}) of the objects.
For Haumea, the relative phasing of data taken at two epochs separated by six months did not pose any problem, thanks
to the highly accurate knowledge of the period\footnote{We re-determined this period as P= 3.915341 $\pm$ 0.000005 h using additional optical data from January 2010 combined with data from 2007 \citep{2010A&A...518L.147L}. A detailed description of these observations and of the technique used to derive this rotation period can be found in \cite{thirouin2013}.}. A running mean of the Haumea thermal fluxes in each filter was applied with a bin of 0.05 in rotational phase (= 11.75 minutes of time), which finally obtained  20 points at green band and 19 at red band. The final thermal 
light curves are shown in Figs. \ref{LC_haumea_g_r}, \ref{LC_vs2blue_red}, and   \ref{LC_az84green}
for Haumea, 2003 VS$_2$ and 2003 AZ$_{84}$, respectively.

%______________________________________________________________

\section{Results and analysis}
\label{Discuss}

\begin{table*}

\caption{Amplitudes of thermal versus optical light curves and estimated thermal shifts} % title of Table

\label{thermal_vs_optical} % is used to refer this table in the text
 
\centering % used for centering table

\begin{tabular}{r c c c c} % centered columns (4 columns)

\hline\hline % inserts double horizontal lines

Object  &       Band    & Thermal $\Delta$m     &       Thermal/Optical & Thermal shift   \\
        &       [$\mu$m]        & [mJy] & amplitude ratio & [minutes of time / degrees / rotational phase]   \\

% table heading

\hline % inserts single horizontal line

(136108) Haumea                 &       100 & 9.8 $\pm$ 0.8 & 1.8 & -2 $\pm$ 1 / -3 $\pm$ 1 / -0.008 $\pm$ 0.003\\   %Error= 0.030 in phase
                                                                &       160 & 7.8 $\pm$ 1.2 & 1.5 & 13 $^{+1}_{-2}$ / 21$^{+1}_{-3}$ / 0.06 $\pm$ 0.01\\ %Error= 0.045 in phase
\hline                                                          
(84922) 2003 VS$_{2}$                   &       70  & 1.7 $\pm$ 1.0 & 0.6 & -4 $\pm$ 18 / -3 $\pm$ 15 / -0.008 $\pm$ 0.041 \\ %Error= 0.13 in phase
                                                                &       160 & 1.8 $\pm$ 1.2 & 0.7 & -24 $\pm$ 45 / -19 $\pm$ 37 / -0.054 $\pm$ 0.102 \\        %Error= 0.13 in phase                                                 
                                                                
\hline %inserts single line

\end{tabular}

\footnotesize{\textbf{Band:} PACS-band in $\mu$m. \textbf{Thermal $\Delta$m:} is the peak-to-peak amplitude (in mJy) derived from the thermal light curve by means of a Fourier fit to the data. \textbf{Thermal/Optical:} is the ratio of the thermal versus the scaled optical amplitudes. \textbf{Thermal shift:} is the time shift of the thermal versus the optical light curve estimated from the Fourier fits of the thermal data and the Fourier fits of the optical light curves. Shifts are expressed in minutes of time, degrees and rotational phase (between 0 and 1).}

\end{table*}

%______________________________________________________________

\subsection{(136108) Haumea}

Haumea's optical light curve is one of the best studied among TNOs \citep{2006ApJ...639.1238R,2007AJ....133.1393L,2008AJ....135.1749L,2010A&A...522A..93T}. Besides its strong amplitude (0.28 magnitudes), its most remarkable feature is its asymmetric character, exhibiting two unequal brightness
maxima, which cannot be explained by a pure shape effect and is interpreted as being due to the presence of a darker (and redder) 
region on the object's surface \citep{2008AJ....135.1749L}.

\begin{figure}[!hpbt]
   \centering
   
   \includegraphics[width=9.5cm,angle=180]{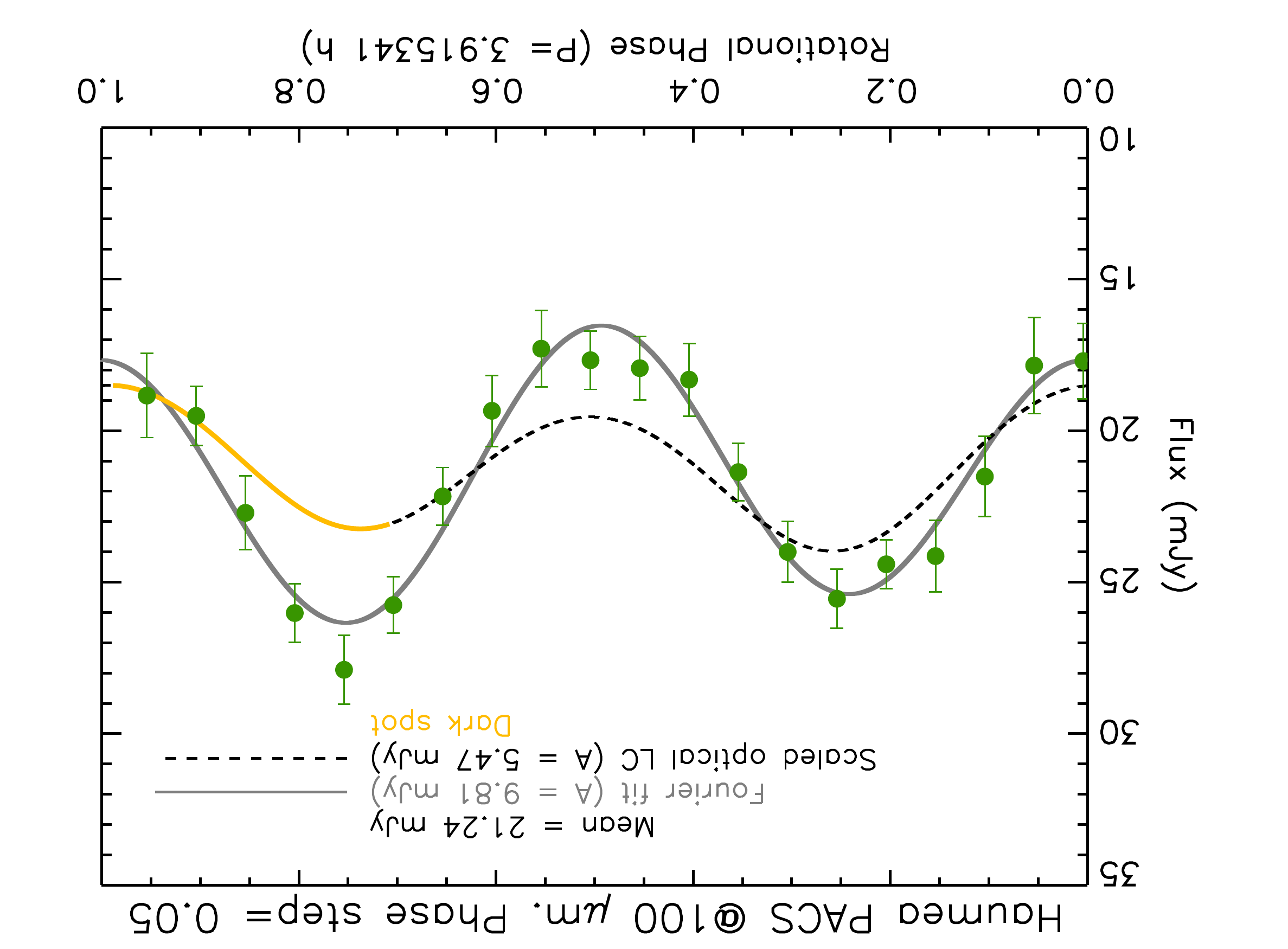}
   \includegraphics[width=9.5cm,angle=180]{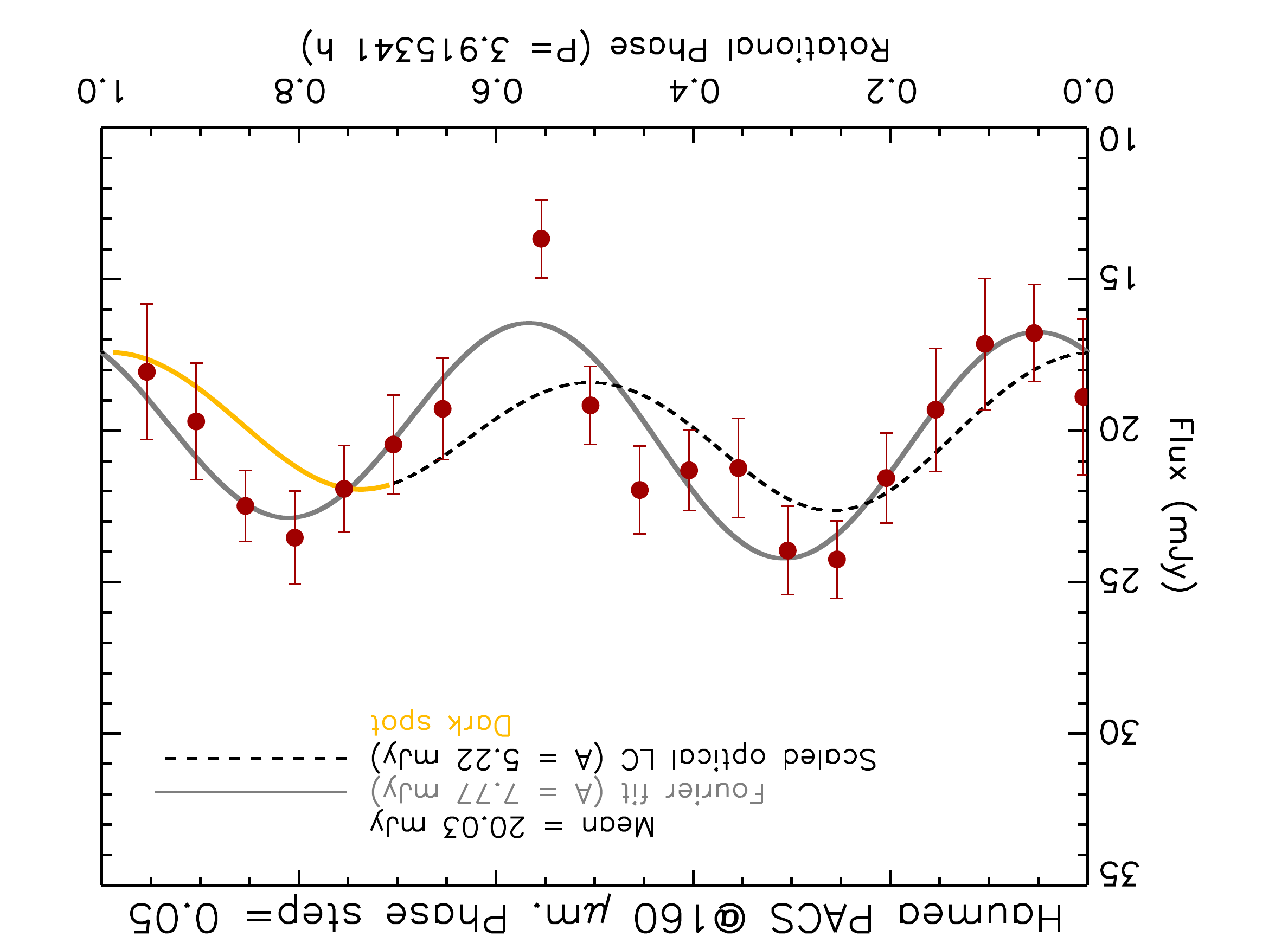}
  
      \caption{Haumea's thermal light curve at 100 $\mu$m (top) and 160 $\mu$m (bottom), combining data
from the two visits (see text for details). The black dashed curve is the ``scaled optical light curve" obtained by
rescaling the optical brightness to match the mean of the thermal fluxes.  
The part of this curve that is affected by Haumea dark spot, according to the \cite{2008AJ....135.1749L} preferred model,
is outlined in yellow. The grey solid curve is a Fourier fit to the thermal data. The reference for zero phase is at JD = 2455188.720000 days, uncorrected for light-time. Rotational phases have been computed using light-time corrected julian dates and light-time corrected zero date (see caption of table \ref{HaumeaPACSdata} for further details). The uncertainty in the rotational phase is $\pm$ 0.001.}
      
         \label{LC_haumea_g_r}

\end{figure}

Figure \ref{LC_haumea_g_r} shows the thermal data for Haumea at 100 and 160 $\mu$m, respectively, as a function of rotational phase,
using a period of 3.915341 h. The zero phase epoch is JD = 2455188.720000 (uncorrected for light-time) and phases
are calculated using light-time corrected julian dates and light-time corrected zero date.
In addition to the thermal fluxes, Fig. \ref{LC_haumea_g_r} displays the Scaled optical LC, which represents the optical fluxes rescaled by some multiplicative factor to match the mean thermal flux level. The part of the optical light curve that is affected by the dark spot is outlined in yellow. The overall positive correlation between the thermal and optical light curves indicate that both are mostly caused by the object elongated shape, as already noted in \cite{2010A&A...518L.147L}, hereafter Paper I. However,
the highest quality 100 $\mu$m data further indicates an asymmetry in the two thermal flux maxima, 
with the strongest occurring near phase $\sim$0.75, i.e. in the part of the optical light curve that is
affected by the dark spot. The possible influence of the spot could not be discerned in  Paper I,
and the present data are of higher quality. On the other hand, the 160 $\mu$m data do not
show evidence for an enhanced thermal flux associated with the dark spot. 

A Fourier fit of the thermal data permits us to determine the amplitude of the thermal light curve (defined
as the difference between maximum and minimum fluxes in the Fourier fit), as well as its phasing relative to
the optical light curve. In both filters, the thermal light curve amplitude is larger than its optical counterpart
and diminishes with increasing wavelength; these behaviors are in accordance with thermophysical model expectations.
However, the two thermal filters do not give fully consistent information of the phase shift between the
thermal and optical data: 100 $\mu$m data appear well in phase with the optical light curve, while
160 $\mu$m data appear shifted by 0.06 in phase (i.e. by about 21 degrees: see Table \ref{thermal_vs_optical}). 

Finally, we perform a consistency check of the fluxes obtained in the thermal light curves. To do this, we run a Near Earth Asteroid Thermal Model \citep[NEATM,][]{Harris98} for the green and red fluxes at the minimum and maximum of the thermal light curves. We assume H$_{V}$ = 0.43 $\pm$ 0.01 mag as in \cite{2013A&A...555A..15F} and the geometric albedo and beaming factor derived in that work (p$_{V}$ = 80.4\%, $\eta$ = 0.95 ). Under these assumptions we estimate the area-equivalent diameter from the NEATM for the minimum of the thermal light curves, using an absolute magnitude of (H$_{V}$ + 0.21/2) mag, where 0.21/2 is the semi-amplitude from the optical light curve, obtaining D$_{min}$ = 1173 km.  Running a NEATM in the same way for the maximum, for an absolute magnitude (H$_{V}$ - 0.21/2) mag, we obtain D$_{max}$ = 1292 km. These diameters are consistent, within error bars, with the best equivalent diameter obtained in \cite{2013A&A...555A..15F}.

%______________________________________________________________

\subsubsection{Haumea modeling} 

\label{haumea-models}

Following our work in \cite{2010A&A...518L.147L}, modeling of the thermal light curve was performed using OASIS \citep[Optimized Astrophysical Simulator for Imaging Systems:][]{2010SPIE.7533E..11J}. OASIS is a versatile tool in which an object is described by triangular facets. The orientation of each facet with respect to pole orientation, Sun direction, observer direction, and time as the object rotates, is calculated. OASIS therefore requires a shape model and an
assumed aspect angle (i.e., pole orientation). For the shape of Haumea, we used an ellipsoid made of 5 120 triangles. For the aspect angle, the large amplitude of Haumea's optical light curve favors a large angle and,  here, we  assumed an equator-on geometry (aspect angle $\theta$ = 90$^{\circ}$). A spectral and bolometric emissivity of 0.9 in all filters is assumed as well.

In Paper I,  two shape models for Haumea  (defined by the a, b, and c semi-major axes of the ellipsoid)  were used, based on optical light curves observations by \cite{2008AJ....135.1749L} and assuming a Jacobi hydrostatic equilibrium figure (a $>$ b $>$ c). The two shape models were derived by considering two different scattering properties for Haumea's surface (Lambertian reflectivity, shape model 1, and Lommel-Seelinger reflectance properties, shape model 2), leading to slightly different values of b/a, c/a, and the object density. For a given shape model, knowledge of the object mass \citep{2009AJ....137.4766R} provided the absolute values of a, b, and c, which in turn provided the object mean geometric albedo, based on its H$_v$ magnitude. All these parameters were then implemented 
in a NEATM thermal model \citep{Harris98}, and the only free parameter in fitting the thermal light curve was the so-called beaming factor, $\eta$. In this process, a phase integral $q$ = 0.7 was adopted; this value is reasonable for a high albedo object (see \citeauthor{2000Icar..147..220L} \citeyear{2000Icar..147..220L}, Fig. 7 and \citeauthor{Brucker2009} \citeyear{2000Icar..147..220L}) but admittedly uncertain. Considering mostly an aspect angle $\theta$ = 90$^{\circ}$, the main conclusion of Paper I was that $\eta$ = 1.15 satisfies the mean thermal flux constraint for both shape models, but matches the light curve amplitude only for model 2, which was therefore favored. The relatively low $\eta$ value (for an object at this distance from the Sun) pointed to a generally low thermal inertia for the surface and significant surface roughness effects \citep[see][]{2013A&A...557A..60L}. Paper I also briefly explored the effect of aspect angle by considering the case $\theta$ = 75$^{\circ}$, and found that such a model could be valid, but using a slightly larger $\eta$ value (e.g. $\eta$ = 1.35 instead of $\eta$ = 1.15). However, the modest quality of the
thermal light curve in Paper I did not warrant the use of more elaborate models.

Given the improved data quality in the current work, including the apparent detection of increased thermal emission at the expected location of the dark spot, we now improve these early models by (i) considering thermophysical models (TPM); (ii) exploring in some detail the effect of a surface spot. The essential physical parameter to constrain is now the surface thermal inertia, $\Gamma$. To make allowances for possible surface roughness effects, however, the TPM can also include an $\eta$ factor, but which in this formalism is by definition $\leq$ 1 (see e.g. \citeauthor{2004A&A...413.1163G} \citeyear{2004A&A...413.1163G}, Eq 3; \citeauthor{2011Icar..214..701L} \citeyear{2011Icar..214..701L}, Eq 2). Unlike NEATM, which for a uniform (constant albedo) elliptical surface, calculates (by construction) a thermal light curve in phase with the optical light curve, the TPM approach enables us to investigate temperature lags owing to thermal inertia. Thus, in principle, the thermal inertia can be derived by investigating the relative phase of the thermal and optical light curves, as constrained by the observations. Once $\Gamma$ is determined, $\eta$ and $q$ may be adjusted so as to match the mean flux level and the amplitude of the light curve. However, the problem may be underconstrained, i.e. $\eta$ and $q$ cannot necessarily be determined separately. If the surface includes a spot of known albedo and spatial extent, the parameters (i.e. $\Gamma$, and $\eta$ and/or $q$) may be adjusted separately in the spot region and outside. In what follows, and given the low thermal inertias we inferred (see below), we found that the observed flux levels did not require to be enhanced by surface roughness, so we simply assumed $\eta$ = 1, recognizing that some degeneracy exists between  $\eta$ and $q$.

\cite{2008AJ....135.1749L} show that the asymmetry in Haumea's optical light curve can be interpreted with different spot models, characterized by the albedo contrast of the spot with respect of its surroundings and its spatial extent. In all cases, the spot is assumed to be centred on Haumea's equator and to lead one of the semi-major axes by 45$^{\circ}$. Possible models range from a very localized (6\% of Haumea maximum cross section) and low albedo spot (30 \% of the non-spot albedo) to much more extended spot (hemispheric) and subdued in contrast (95~\% of non-spot albedo). In a brief study of the spot effect on the thermal light curve, Paper I considered a spot covering 1/4 of Haumea's maximum projected cross section, with an albedo contrast (about 80 \%) of the non-spot albedo, as prescribed by the \cite{2008AJ....135.1749L} results. The same spot description was adopted here. Being relatively limited in extent, the spot has negligible effect on about half of the thermal light curve (from phase $\sim$ 0.0 to $\sim$ 0.5 with the adopted phase convention). Therefore, as a first step, we focus on the part of the light curve that is not affected by the spot. 

\begin{figure}[hpbt]
   \centering
   
  \includegraphics[width=9.5cm]{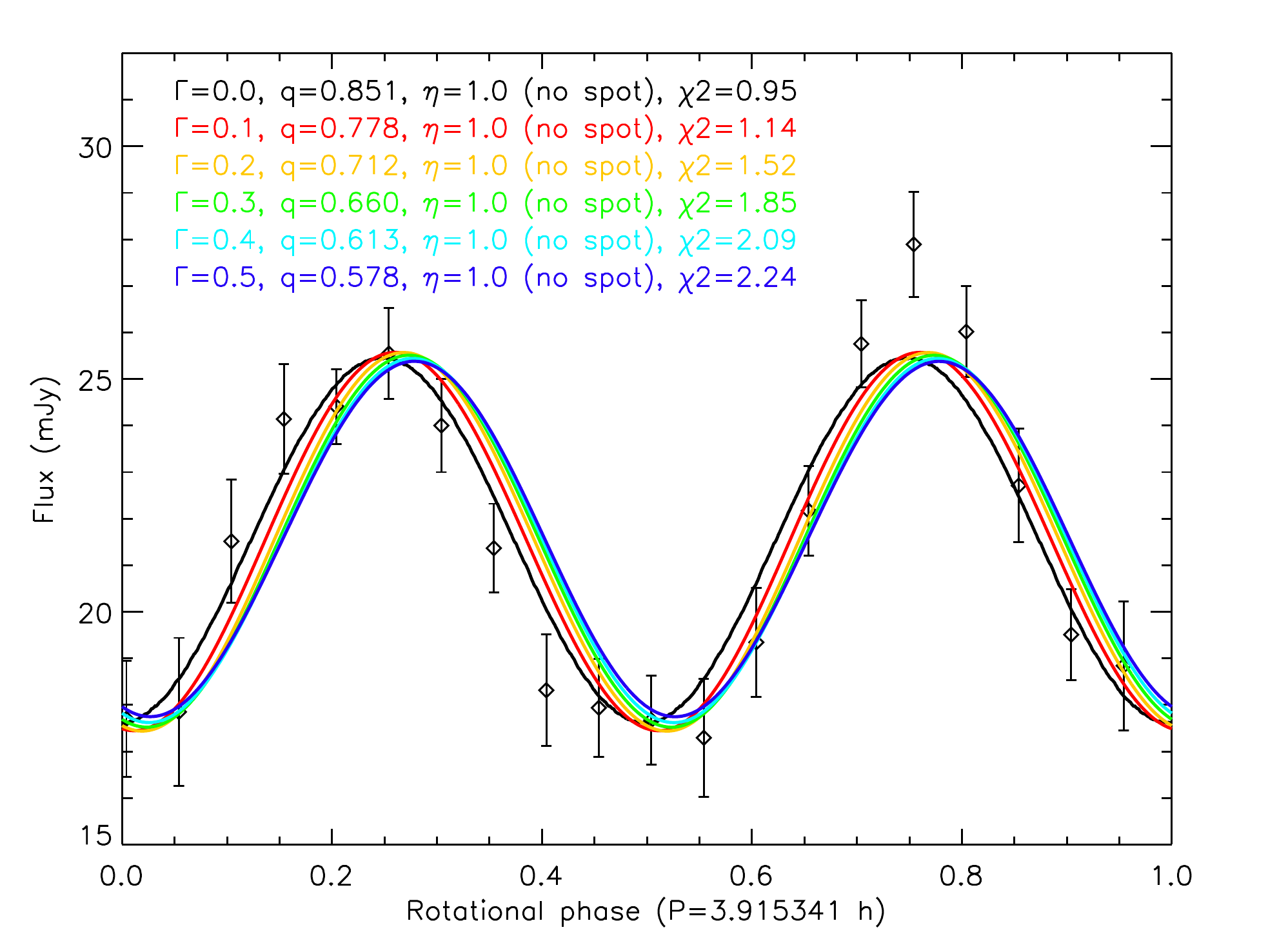}
  
      \caption{Comparison of the Haumea 100 $\mu$m light curve with homogeneous models (no spot) having various
values of the thermal inertia $\Gamma$ in the range 0.0 to 0.5 MKS. No surface roughness effects are included
(i.e. $\eta$ = 1). For each $\Gamma$ value, the phase integral is adjusted to match the mean flux level and
light curve amplitude. The lowest $\Gamma$ values (0.0-0.2 MKS) provide the best fit to the data. }
       \label{figXX}
\end{figure}

Fig. \ref{figXX} shows the comparison of the observed 100 $\mu$m light curve with several homogeneous (no spot) models differing by their surface thermal inertia ($\Gamma$ = 0.0 to 0.5 MKS by steps of 0.1). Here, and throughout the following, shape model 2 is adopted \citep[this shape model is also favored by the analysis of][]{2014EM&P..tmp....3L} with an aspect angle $\theta$ = 90$^{\circ}$, giving a = 961 km, b = 768 km, c = 499 km, and p$_v$ = 0.71. These a, b, c values lead to a mean area-equivalent diameter D$_{equiv}$= 2$\cdot$ a$^{1/4}\cdot$ b$^{1/4}\cdot$ c$^{1/2}$ = 1309 km, within the error bars of the area-equivalent diameters obtained from radiometric techniques for this object (1324$\pm$167 km from \citeauthor{2010A&A...518L.147L} \citeyear{2010A&A...518L.147L}; 1240$^{+69}_{-58}$ km from \citeauthor{2013A&A...555A..15F} \citeyear{2013A&A...555A..15F}). Using a shape model that is consistent with the optical light curve is preferable to a radiometric solution that may include measurements at different light curve phases. For each value of $\Gamma$, the phase integral ($q$) is adjusted to provide the best fit to the mean flux level and amplitude of the light curve outside the spot region (i.e. at phases 0.0--0.5). For $\Gamma$ = 0.0, 0.1, 0.2, 0.3, 0.4, and 0.5 MKS, respectively, the required values of $q$ are 0.851, 0.778, 0.712, 0.660, 0.613, and 0.578, respectively. As can be seen in Fig. \ref{figXX}, $\Gamma$ = 0.0 provides the best fit to the part of the light curve not affected by the spot, while larger $\Gamma$ values progressively lead to larger delays of the thermal emission, which are not observed in the 100 $\mu$m data. Detailed comparisons of the data show that the reduced $\chi^2$ is minimum for $\Gamma$ = 0.0 MKS and larger but still reasonable for $\Gamma$ = 0.2 MKS. From this we conclude that $\Gamma$ values in the range 0.0--0.2 MKS are consistent with the data. The associated phase integral values are in the range 0.851--0.712.
These are generally consistent but still somewhat larger than the 0.7 value assumed in Paper I. We note that using the \cite{Brucker2009} empirical relationship between geometric albedo and phase integral, phase integrals of 0.851 (obtained for $\Gamma$ = 0 MKS) -- 0.712 (for $\Gamma$ = 0.2 MKS) would imply geometric albedos in the range 1.05--0.64. The latter value is more consistent with $p_{v}$= 0.71, as indicated by shape model 2. From this point of view, $\Gamma$ = 0.1--0.2 MKS would seem a more plausible solution but, as indicated above, the associated fits are worse than with $\Gamma$ = 0 MKS. We also note that these $q$ values hold for our assumption of $\eta$ = 1 (no surface roughness), and that for a given $\Gamma$, any finite surface roughness would require an even larger value of $q$.

\begin{figure}[hpbt]
   \centering
   
   \includegraphics[width=9.5cm]{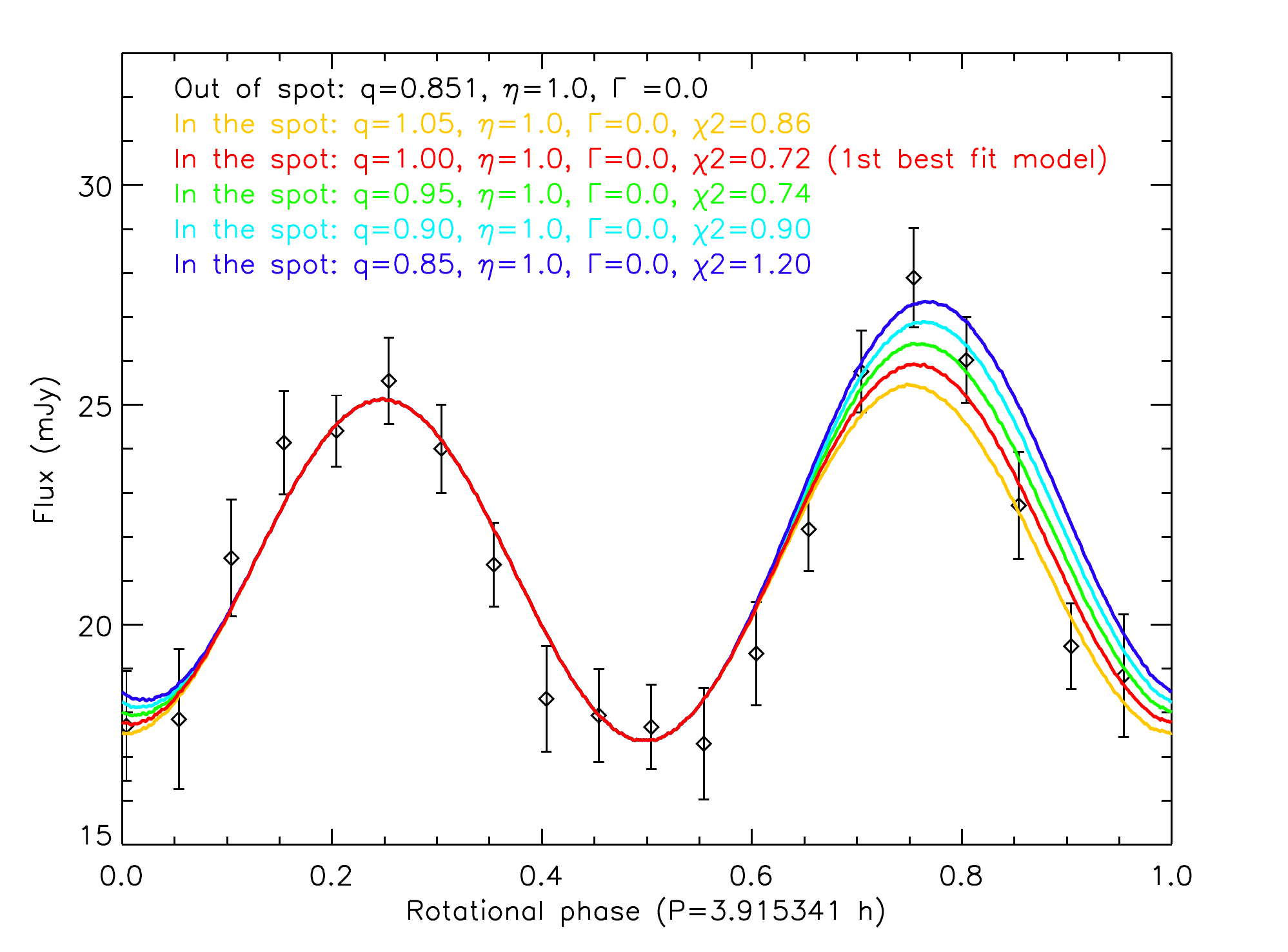}
  
      \caption{Models with variable phase integral in the spot region. In these models, the spot
thermal inertia is $\Gamma_{spot}$ = 0, as in the no-spot region.  The red curve (having $q$ = 1.00 in the spot) 
represents the first best fit model for the 100 $\mu$m data.}

         \label{figYY}
         
\end{figure}

We now turn to constraints on the dark spot. Continuing with the assumption of no surface roughness, we investigate the effect of a specific phase integral or thermal inertia in the spot. In all models, the spot has an equivalent geometric albedo of 79\% of its surroundings (p$_v$ = 0.71), i.e. $p_v$= 0.56. Figure \ref{figYY} shows the effect of changing the phase integral in the spot ($q_{spot}$) by steps of 0.05, maintaining its thermal inertia to $\Gamma_{spot}$ = 0 as in the best fit solution of the no-spot region (Fig. \ref{figXX}). As is clear from Fig. \ref{figYY}, mild changes in $q_{spot}$, e.g. from $\sim$0.85 to $\sim$1.05 produce flux variations at the adequate level.  The flux excess near phase 0.75 would point to $q_{spot}$ = 0.85, essentially identical to the non-spot region (meaning that the flux excess is purely an effect of the spot darker albedo). However this model (blue curve in Fig. \ref{figYY}) clearly overpredicts the fluxes beyond the peak phase, where data are best fit with $q_{spot}$ = 0.95--1.05. The best overall fit of the light curve using an $\chi^2$ criterion is achieved with $q_{spot}$ = 1.00, and the model light curve shows only a very weak flux excess in the spot region (meaning that the effect of the darker spot albedo is essentially compensated for by the effect of a larger phase integral). A phase integral of 1.00 seems very high (and not in line with the \citeauthor{Brucker2009} \citeyear{Brucker2009} relationship), but it has been observed in other solar system objects \citep[see][for Triton]{1991JGR....9619203H}. However, while the fit is satisfactory (first best-fit model in Fig. \ref{figYY}), we note that it might not be a very realistic solution, since there is a general positive correlation between albedo and phase integral on airless bodies (\citeauthor{2000Icar..147..220L} \citeyear{2000Icar..147..220L}, Fig. 7; \citeauthor{Brucker2009} \citeyear{Brucker2009}), in contradiction with the dark (low albedo) spot, which has a larger phase integral (higher albedo) than its surroundings.

\begin{figure}[hpbt]
   \centering
   
   \includegraphics[width=9.5cm]{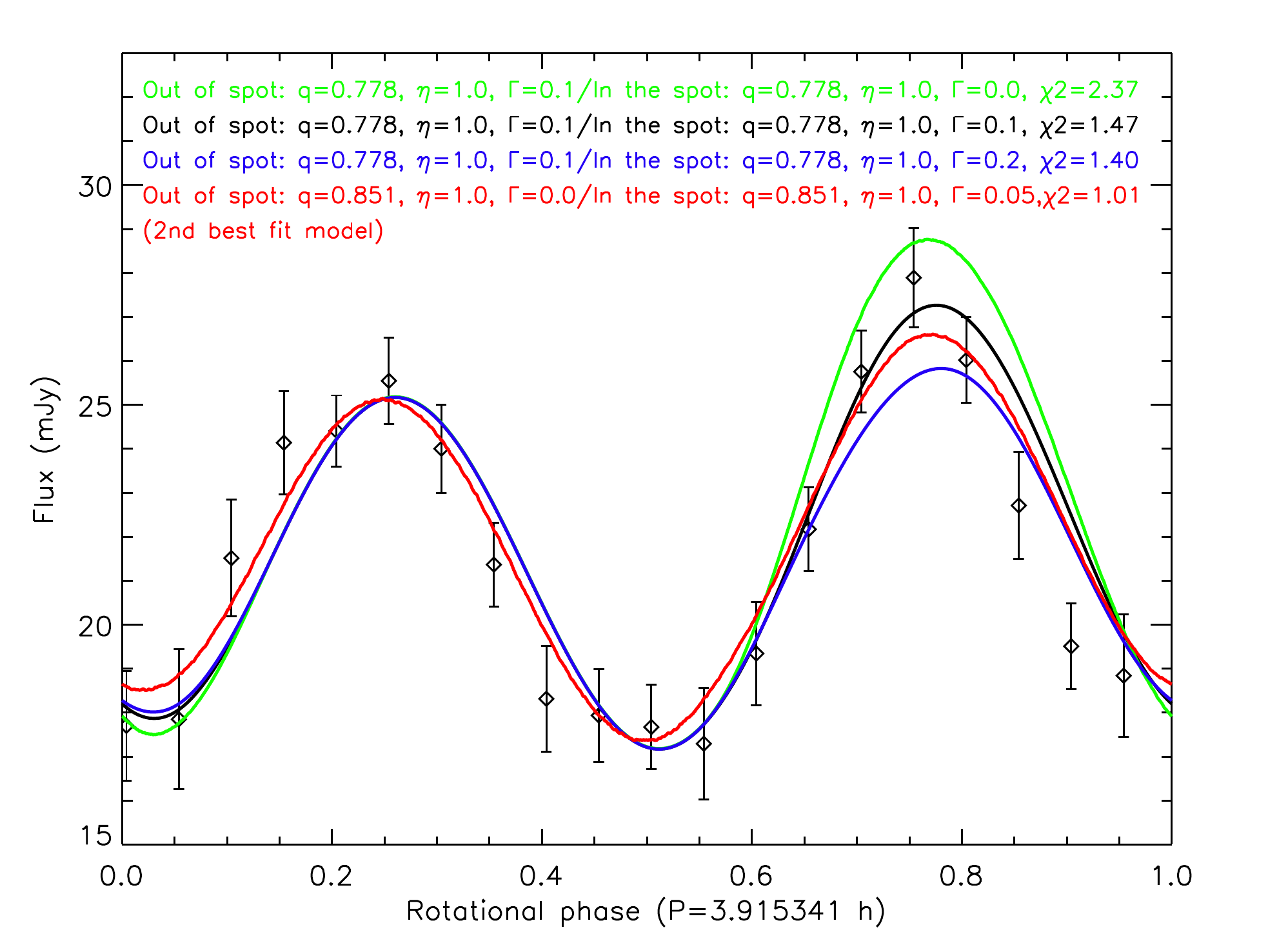}
  
      \caption{Models with variable thermal inertia in the spot region. In these models, the spot phase integral is kept at the same value as in the no-spot region. The green, black, and blue curves are for $\Gamma$ = 0.1 MKS and $q$ = 0.778 outside of the spot,  and $\Gamma$ = 0.0, 0.1, and 0.2 MKS in the spot.
The red curve represents the second best fit model at 100 $\mu$m with $\Gamma$ = 0.0 MKS and $q$ = 0.851 outside of the spot, and  $\Gamma_{spot}$ = 0.05 MKS  in the spot.}
      
         \label{figZZ}
\end{figure}

Figure \ref{figZZ} shows the effect of changing the spot thermal inertia, maintaining its phase integral to the value of its surroundings. For this exercise, we first consider $\Gamma$ = 0.1 MKS in the no-spot region, associated with $q$= 0.778. We maintain the $q_{spot}$ at this value and show models with $\Gamma_{spot}$ = 0.0, 0.1, and 0.2 MKS. These models show that variations of $\Gamma$ within this range produce flux changes at the appropriate level. In this family of models, the best fit is achieved with $\Gamma_{spot}$ = 0.2 MKS (blue curve in the Fig.), which again suggest that, to avoid too large fluxes past the peak, the thermal effect of darker albedo needs to be partly compensated for by a larger thermal inertia. As discussed above, however, $\Gamma$ = 0.1 MKS is not an optimum fit of the no-spot part of the light curve. The red curve in Fig. \ref{figZZ} shows a model with $\Gamma$ = 0.0 MKS and $q$ = 0.851 outside of the spot, and  $\Gamma_{spot}$ = 0.05 MKS (and still $q_{spot}$ = 0.851) in the spot (second best fit model). This model fit is worse than the first best fit model ($\Gamma_{spot}$ = 0.0 MKS, $q_{spot}$ = 1.00) and, in fact, no better than the no-spot model: the fit improves in the region of flux maximum around phase 0.75, but the non-zero thermal inertia in the spot tends to delay the model too much in the 0.85--1.0 phase region. Thus the models of Fig. \ref{figZZ} indicate that, at most, the thermal inertia in the spot region is slightly higher than in its surroundings.
This type of behaviour might be expected. A statistical study of TNO thermal properties  \citep{2013A&A...557A..60L} suggests that the highest albedo TNOs generally exhibit particularly low thermal inertia. A plausible explanation is that in addition to a specific composition (e.g. pure ice), a higher albedo may reflect a smaller grain size. In fact, the generally low thermal inertia of TNOs points to a regime where radiative conductivity (i.e. in surface pores) is important in the overall heat transfer, thus high albedo objects might plausibly be associated with low thermal inertia, and the association of lower albedo with larger thermal inertia is also plausible. 

\begin{figure}[hpbt]
   \centering
   
   \includegraphics[width=9.5cm]{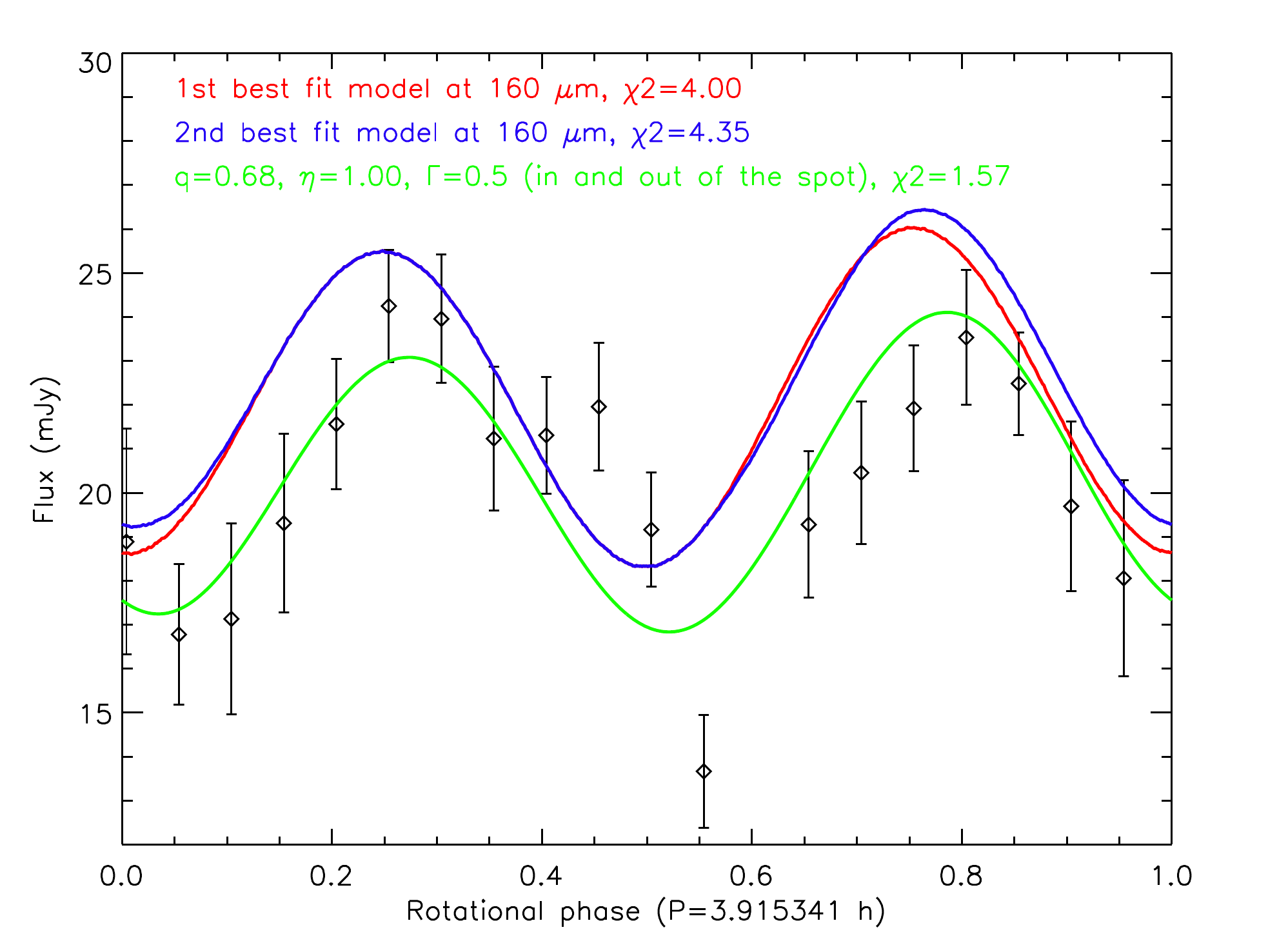}
  
      \caption{The two best fit models from the 100 $\mu$m light curve (see Figs. \ref{figYY} and  \ref{figZZ}) applied to the 160 $\mu$m light curve (red and blue curves, respectively). The green curve is the best-fit model in terms of least squares adjusted to match the 160 $\mu$m phase, having $\Gamma$= 0.5 MKS and $q$ = 0.68 throughout Haumea's surface.}
      
         \label{figTT}
         
\end{figure}

Fig. \ref{figTT} shows the application of the above two best fit models to the Haumea's 160 $\mu$m light curve. In both cases, the agreement with the observed 160 $\mu$m mean flux, and especially light curve amplitude is reasonable, but the models fail to match the data in two respects: (i) the modeled flux levels are on average too high (by $\sim$ 10\%), (ii) the models are somewhat out of phase with the observations. The first problem may suggest a calibration error in the 160 $\mu$m data (more subject to sky contamination). The second one is related to the fact that the 160 $\mu$m data appear to be shifted by about 0.06 $\pm$ 0.01 in rotational phase ($\sim$ 21 degrees or $\sim$ 13 minutes of time, see Table \ref{thermal_vs_optical}) with respect to the 100 $\mu$m data. Optimum phasing of the model with respect to the 160 $\mu$m data would require a thermal inertia $\Gamma$ $\sim$ 0.5 MKS, and fitting the mean flux levels would then require a phase integral q = 0.68. This model tailored to the 160 $\mu$m data does not include any specific values of $\Gamma$ or q in the dark spot region, as this is not required by these data. Overall a simultaneous fit of the 100 and 160 $\mu$m data within error bars does not appear possible. Although  so-called compromise parameters could be formally found by performing an $\chi^{2}$ minimization on both datasets simultaneously, we feel that separate modeling is preferable since it provides a handle on model limitations and realistic range of solution parameters. Giving more weight to the higher quality 100 $\mu$m data, we favor $\Gamma$ = 0.0--0.2 MKS but, based on 160 $\mu$m modeling, we regard $\Gamma$ = 0.5 MKS as an upper limit to Haumea's thermal inertia.\\

In summary, thermophysical modeling of the Haumea light curves indicates that: (i) the object's thermal inertia $\Gamma$ is extremely small (less than 0.5 MKS and probably less than 0.2 MKS) and its phase integral is high (at least 0.8 for $\Gamma$ = 0.1 MKS and probably even higher if surface roughness is important); (ii) only small changes in the surface properties of the dark spot (e.g. changes in the thermal inertia by $\sim$ 0.1 MKS or in the phase integral by $\sim$ 0.1) are required to significantly affect the emitted fluxes on the hemisphere where the spot resides; larger changes are excluded by the data; (iii) the most plausible scenario may invoke a slightly higher thermal inertia in the dark spot compared to its surroundings, but a fully consistent picture is still not found, since the $\sim$ 21 degrees shift in phase between the 100 and 160 $\mu$m data is difficult to understand.

Finally, we note that in all of our Haumea models we ignored the possible contribution of its satellites Hi'iaka and Namaka. Although their albedos are unknown,
these moons are thought to have been formed by a catastrophic impact that excavated them from the proto-Haumea ice mantle
and led to the Haumea family \citep{2007Natur.446..294B} or from rotational fission \citep{2012MNRAS.419.2315O}. As such, their albedos are probably comparable to, or even higher than,
Haumea's itself. Assuming a 0.70 geometric albedo, Hi'iaka and Namaka diameters are $\sim$320 and $\sim$160 km, respectively
\citep{2016AJ....151..148T}. Furthermore, assuming identical thermophysical properties, they would contribute in proportion of their projected surfaces, i.e. 6 \% and 1.5 \% of Haumea's thermal flux. Although this is not negligible, we did not include this contribution
owing to its uncertain character. Should the Haumea's thermal fluxes to be modeled decrease by $\sim$20 mJy x 7.5 \% $\sim$ 1.5 mJy,
this could be taken care of in the models by a slight increase of the phase integral for a given thermal inertia, without
any changes to the conclusions on the object thermal inertia and the dark spot properties.

%______________________________________________________________

\subsection{(84922) 2003 VS$_{2}$}

2003 VS$_{2}$ is a Plutino without known satellites. Near infrared spectra of this body shows the presence of exposed water ice \citep{2008AJ....135...55B}, which probably increases the geometric albedo of this Plutino \citep[$\sim$ 15\%, according with][]{2012A&A...541A..93M}, compared with the mean albedo of TNOs without water ice. This object presents an optical light curve with moderately large peak-to-peak amplitude $\sim$ 0.21 $\pm$ 0.01 mag and a double-peaked rotational period $\sim$ 7.42 hours \citep{2006A&A...447.1131O,2007AJ....134..787S,2010A&A...522A..93T}. To fold the Herschel data to the rotation period with enough precision, we refined the knowledge of the rotational period. To achieve this goal, we used optical observations taken on 4-8 September 2010 by means of the 1.5-m telescope at Sierra Nevada Observatory (OSN, Spain) using a 2k $\times$ 2k CCD with a FOV of 7.8$\arcmin$ $\times$ 7.8$\arcmin$ and 2 $\times$ 2 binning mode (image scale = 0.46$\arcsec$/pixel). Then we merged these observations with old optical light curves obtained also at OSN on 22, 26, 28 December 2003 and 4, 19-22 January 2004, and with 2003 observations published in \cite{2007AJ....134..787S}. No filter was used to perform the OSN observations. We reduced and analysed all these data as  described in \cite{2010A&A...522A..93T} to finally obtain an accurate rotational period of 7.4175285 $\pm$ 0.00001 h ($\Delta$m = 0.21 $\pm$ 0.01 mag.). A further description of the observations and techniques leading to this very accurate rotational period are detailed in \cite{thirouin2013}.

\begin{figure}[!hpbt]
   \centering
   
   \includegraphics[width=9cm]{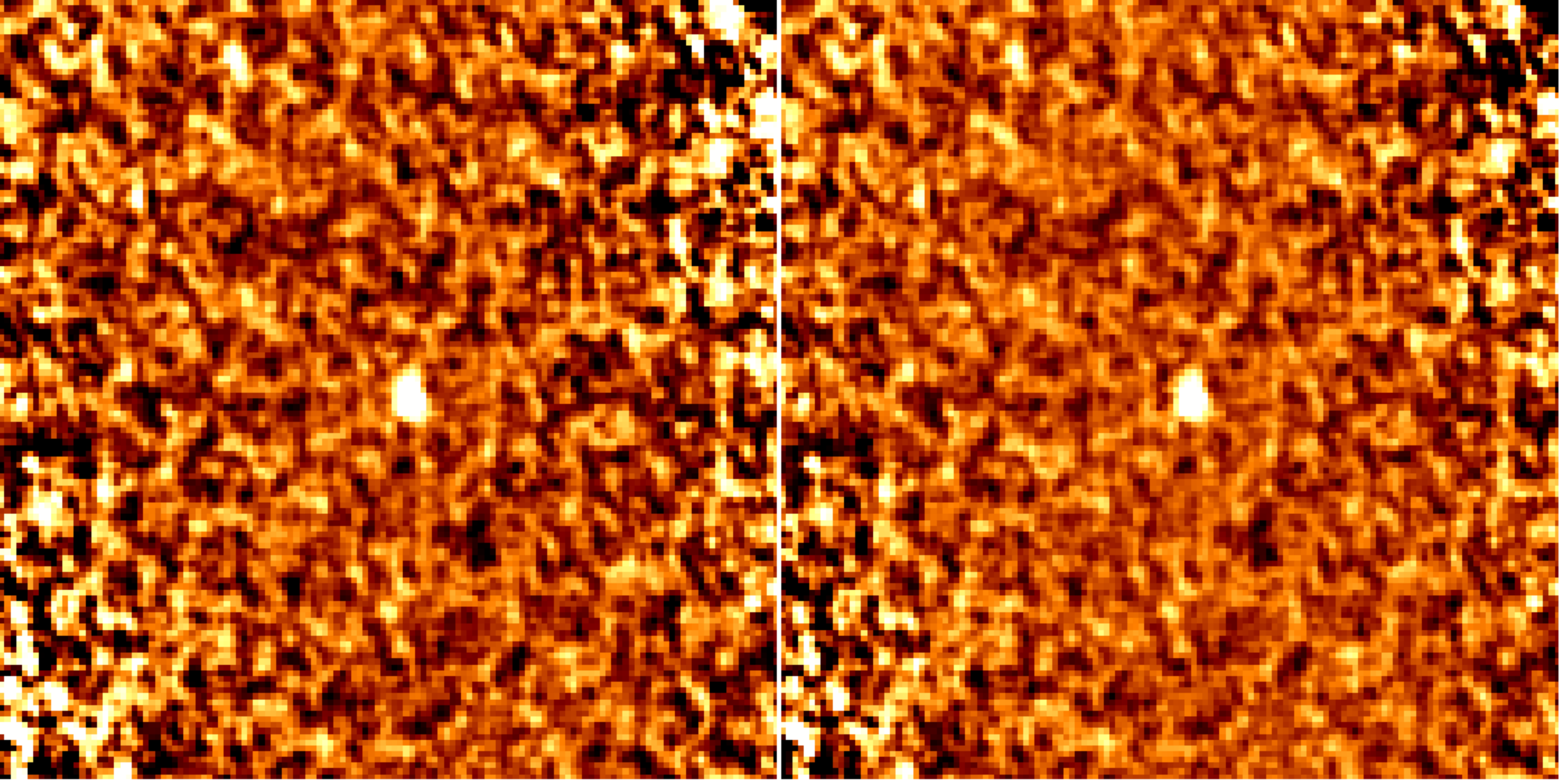}
   \includegraphics[width=9cm]{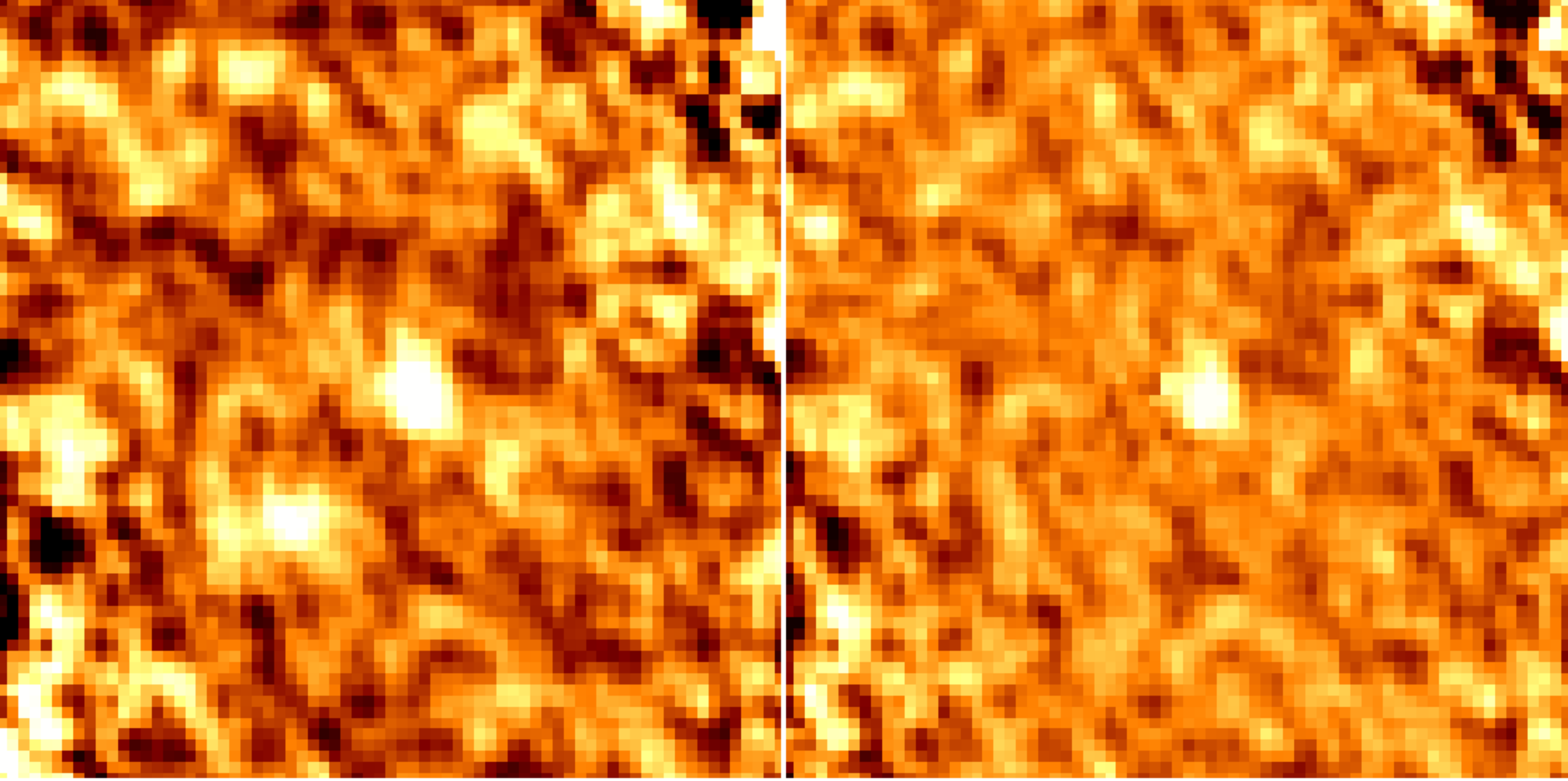}
  
     \caption{Images (top line: 70 $\mu$m; bottom line: 160 $\mu$m) extracted from the thermal time series curve of 2003 VS$_{2}$. Left: original images. Right: background-subtracted images. The FOV is 2.5$\arcmin$ x 2.5$\arcmin$. 2003 VS$_{2}$ is at centre. }
       
         \label{2003VS2_path}

\end{figure}

\begin{figure}[!hpbt]
   \centering
   
   \includegraphics[width=9.5cm,angle=180]{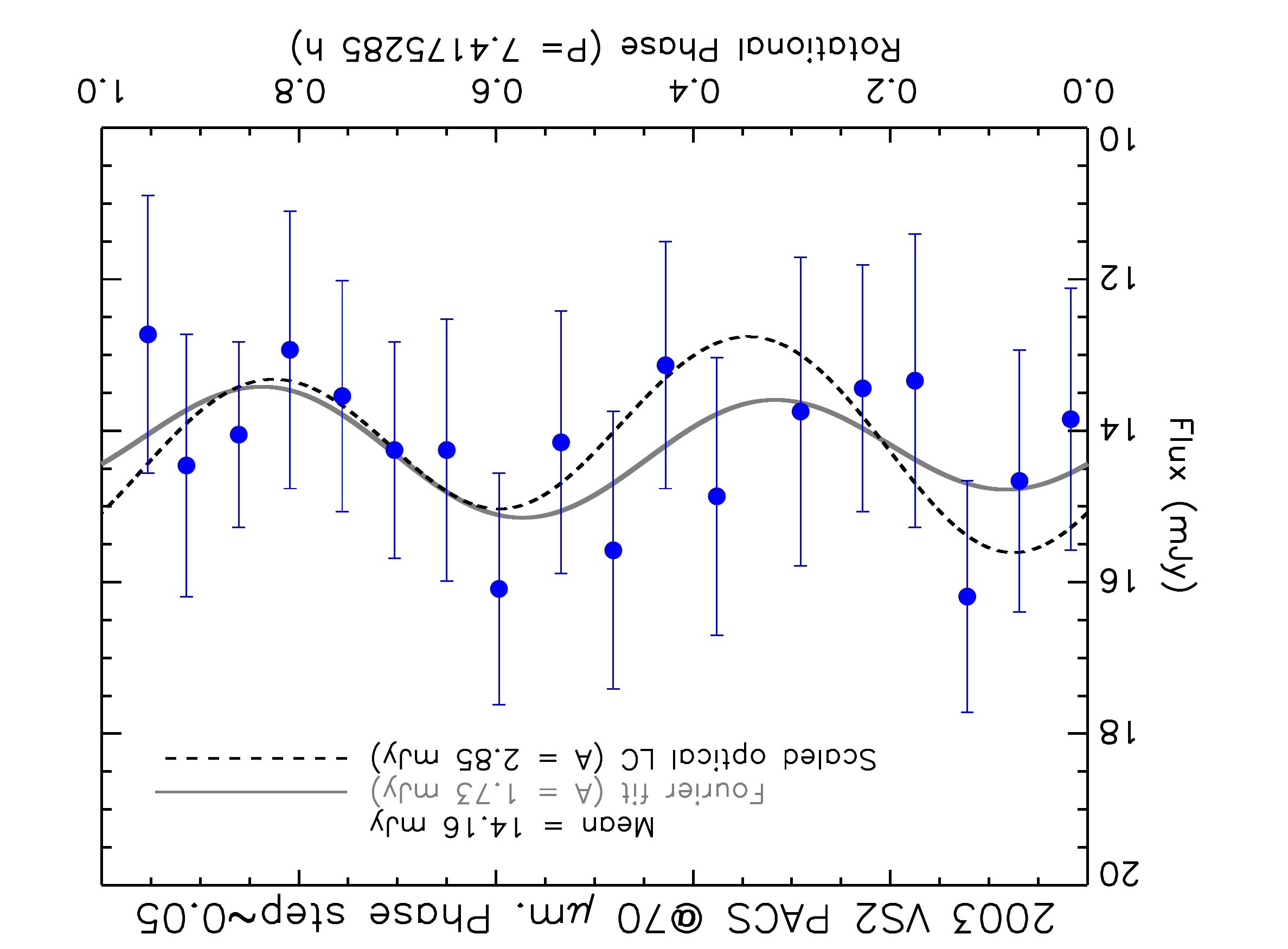}
   \includegraphics[width=9.5cm,angle=180]{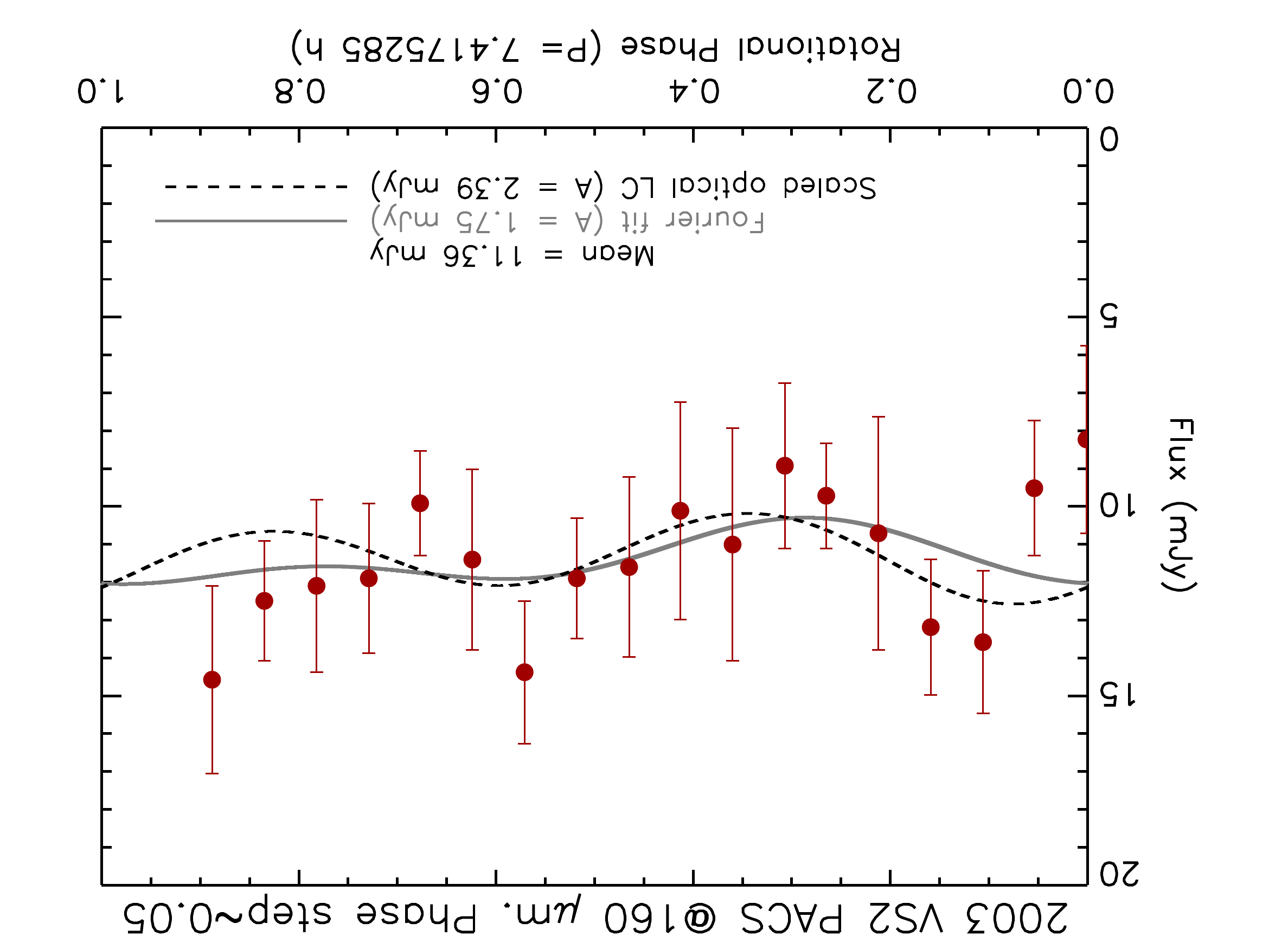}

      \caption{
Thermal time series curve at 70 $\mu$m (top) and 160 $\mu$m (bottom) for 2003 VS$_{2}$. The black dashed curve is the scaled optical light curve obtained by rescaling the optical brightness to match the mean of the thermal fluxes.  
 The grey solid curve is a Fourier fit to the thermal data. The reference for zero phase is at JD = 2452992.768380 days, uncorrected for light-time. Rotational phases have been computed using light-time corrected julian dates and light-time corrected zero date (see caption of Table \ref{VS2PACSdata} for further details). The uncertainty in the rotational phase is $\pm$ 0.01. 
 }              
         \label{LC_vs2blue_red}
         
\end{figure}

The thermal light curves at 70 $\mu$m and 160 $\mu$m are not firmly detected with only a 1.7 $\sigma$ and 1.5 $\sigma$ confidence levels respectively (see Table \ref{thermal_vs_optical} and Fig. \ref{LC_vs2blue_red}). A Fourier fit of the 70 $\mu$m data indicates a mean flux of 14.16 mJy and an amplitude of 1.73 mJy, which is slightly smaller than the optical light curve amplitude. The same analysis yields a negligible shift in time of the 70 $\mu$m data relative to the optical (Fig. \ref{LC_vs2blue_red}). The lower quality 160 $\mu$m
data would suggest a -0.054 $\pm$ 0.102 phase shift (see Table \ref{thermal_vs_optical}). We do not consider this last negative shift significant since it is well below the estimated error bars. In what follows, we pursue with our thermophysical modeling,
focusing on the mean thermal flux and light curve amplitude at 70 $\mu$m.

%______________________________________________________________

\subsubsection{2003 VS$_{2}$ modeling} 
\label{VS2_TPM}

The large amplitude of the optical light curve and its double peaked nature indicates that the main cause for the variability is a triaxial shape \citep{2007AJ....134..787S,2010A&A...522A..93T}. Then, if we assume that the optical light curve is entirely shape-driven, its period and amplitude can be used to derive a shape model under the assumption of hydrostatic equilibrium.
As for Haumea, the large object size ($\sim$500 km) makes the assumption of a Jacobi hydrostatic equilibrium figure (semi-major axes
$a> b> c$) reasonable. Further assuming an equator-on viewing geometry (aspect angle = 90$^{\circ}$), the $\Delta m =$ 0.21 mag
light curve amplitude is related to shape by

\begin{equation} \Delta m = 2.5 \cdot log \left( \frac{a}{b} \right)
\label{deltamred}
.\end{equation}

\begin{figure}[!hpbt]
   \centering
   
   \includegraphics[width=9.5cm,angle=0]{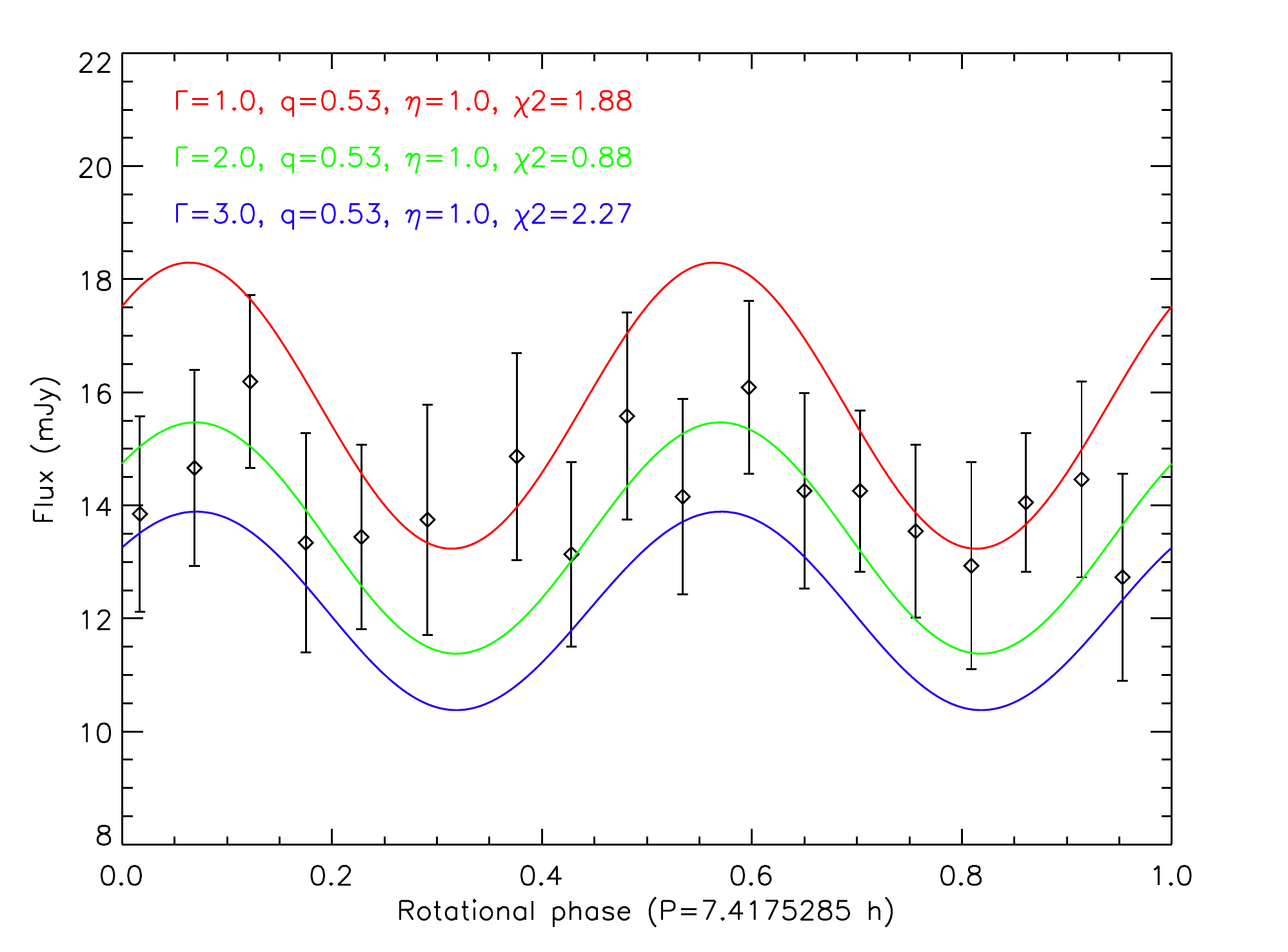}
  
      \caption{Thermophysical model fits of the 2003 VS$_2$ 70 $\mu$m data for various values of the 
thermal inertia. From top to bottom: $\Gamma$ = 1.0, 2.0 and 3.0 MKS. Shape model and other physical input parameters
are given in the text.}
      
         \label{2003VS2_models_blue}
         
\end{figure}

\begin{figure}[!hpbt]
   \centering
   
   \includegraphics[width=9cm]{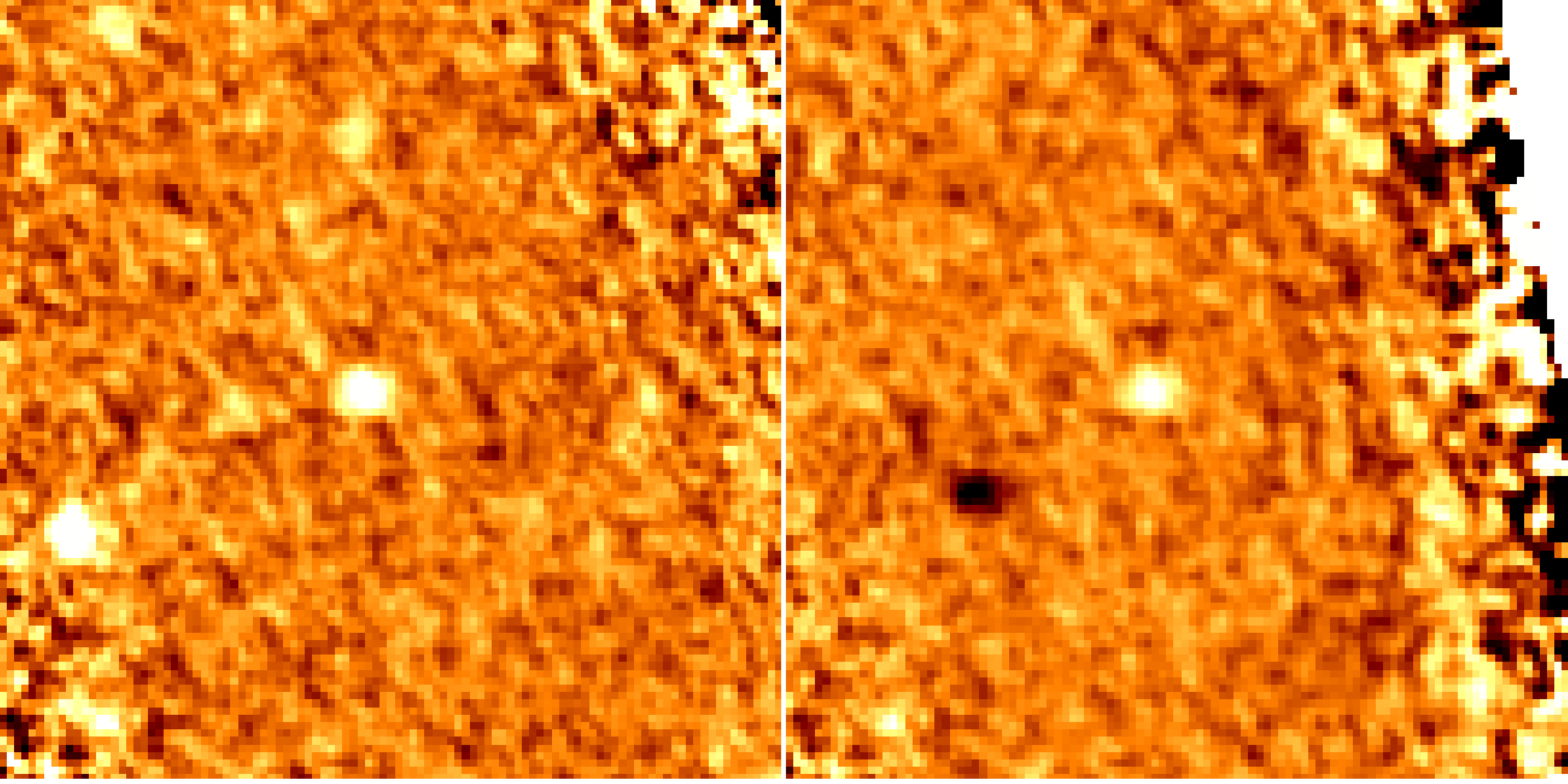}

     \caption{Left panel: one of the 2003 AZ$_{84}$ images at 100 $\mu$m (green) with the original background. Right panel is the same map after the application of the double-differential background subtraction technique, as described in \cite{2013ExA...tmp...36K}. 2003 AZ$_{84}$ is the source at the center of the images. In the right panel, the black source at the bottom left of 2003 AZ$_{84}$ is the negative image of 2003 AZ$_{84}$ resulting from the application of this background-removing technique. The FOV of each image is 2.5$\arcmin$ x 2.5$\arcmin$.  }
       
         \label{2003AZ84_path}

\end{figure}

Using the rotation period of 7.42 h and the Chandrasekhar figures of equilibrium tables for the Jacobi ellipsoids \citep{Chandra87}, we obtain $b/a = 0.82$, $c/a = 0.53$ and $\rho$ = 716 $kg/m^{3}$, where $\rho$ is a lower limit to the density because, if the object is not observed equator-on, the true a, b axial ratios may be higher and the implied density would be also higher. Using the area-equivalent radiometric diameter (D$_{equiv}$ = 523 km) derived from earlier thermal modeling of Herschel and Spitzer data \citep{2012A&A...541A..93M,2013A&A...557A..60L}, defined as D$_{equiv}$= 2$\cdot$ a$^{1/4}\cdot$ b$^{1/4}\cdot$ c$^{1/2}$, we obtain the values of the semi-major axes of 2003 VS$_{2}$: a= 377 km, b= 310 km, and c= 200 km. We further adopt a phase integral q = 0.53 and a V geometric albedo p$_{V}$ = 0.147 from the previous papers and, as for Haumea, we do not consider surface roughness (i.e. $\eta$ = 1). All these values
are used as input parameters to the OASIS code for modeling the 70 $\mu$m thermal light curve of 2003 VS$_{2}$. With this approach, the only free parameter is the thermal inertia $\Gamma$. Figure \ref{2003VS2_models_blue} shows model results for 
$\Gamma$ = 1.0, 2.0 and 3.0 MKS. In this figure, the phase of the thermal models is determined by
requiring that the model with $\Gamma$ = 1.0 matches the observed phase of the thermal data (i.e. a maximum at phase 0.55, see top panel of Fig.\ref{LC_vs2blue_red}). We note that, unlike in the Haumea case (see Fig. \ref{figXX}), models with the various
thermal inertias all appear to be approximately in phase. The difference in behaviour is caused by the combination of the
longest period and warmer temperatures at 2003 VS$_2$ versus Haumea. For a given thermal inertia, this causes a much smaller value of the thermal parameter for 2003 VS$_2$ \citep[see e.g.][]{2013A&A...557A..60L}. A thermal inertia of 2.0 MKS matches reasonably well
the mean flux levels and the light curve amplitude, while the other two models significantly over- or underestimate
the mean flux. Thus, we conclude to a thermal inertia $\Gamma$ = (2.0 $\pm$ 0.5) MKS for 2003 VS$_2$. This value
is fully consistent with the mean thermal inertia for TNOs and centaurs derived statistically from the Herschel TNOs are Cool sample \citep[$\Gamma =$ 2.5 $\pm$ 0.5 MKS,][]{2013A&A...557A..60L}, but significantly above that for Haumea. As indicated in the above paper,
high-albedo objects seem to have preferentially low thermal inertias.

%______________________________________________________________

\subsection{(208996) 2003 AZ$_{84}$}
\label{az84}

2003 AZ$_{84}$ is a binary Plutino with a shallow optical light curve ($\Delta$m $\sim$ 0.07 mag) and with a rotational period $\sim$ 6.79 hours, assuming a single-peaked light curve \citep{2003EM&P...92..207S,2006A&A...447.1131O,2010A&A...522A..93T}. Nonetheless, the double-peaked solution, which corresponds to a rotational period $\sim$ 13.58 hours, cannot be totally discarded \citep{2010A&A...522A..93T}. Near infrared spectra have detected water ice on its surface \citep{2008AJ....135...55B,2009Icar..201..272G,2011Icar..214..297B}. 

As for the other two TNOs, we acquired additional time series images of 2003 AZ$_{84}$ on 4-5 February 2011 with the 1.23-m telescope at Calar Alto Observatory (CAHA, Spain), equipped with a 2k $\times$ 2k CCD camera in the R filter. These data are merged with old CAHA and OSN data from 2003 and 2004 to refine the rotational period, obtaining P = 6.7874 $\pm$ 0.0002 h ($\Delta$m = 0.07 $\pm$ 0.01 mag), which we nominally use to fold the Herschel/PACS data and compare them with the visible light curve. A more detailed description of the observations and techniques of analysis leading to this rotation period are included in \cite{thirouin2013}.

\begin{figure}[!hpbt]
   \centering
   
   \includegraphics[width=9.5cm,angle=180]{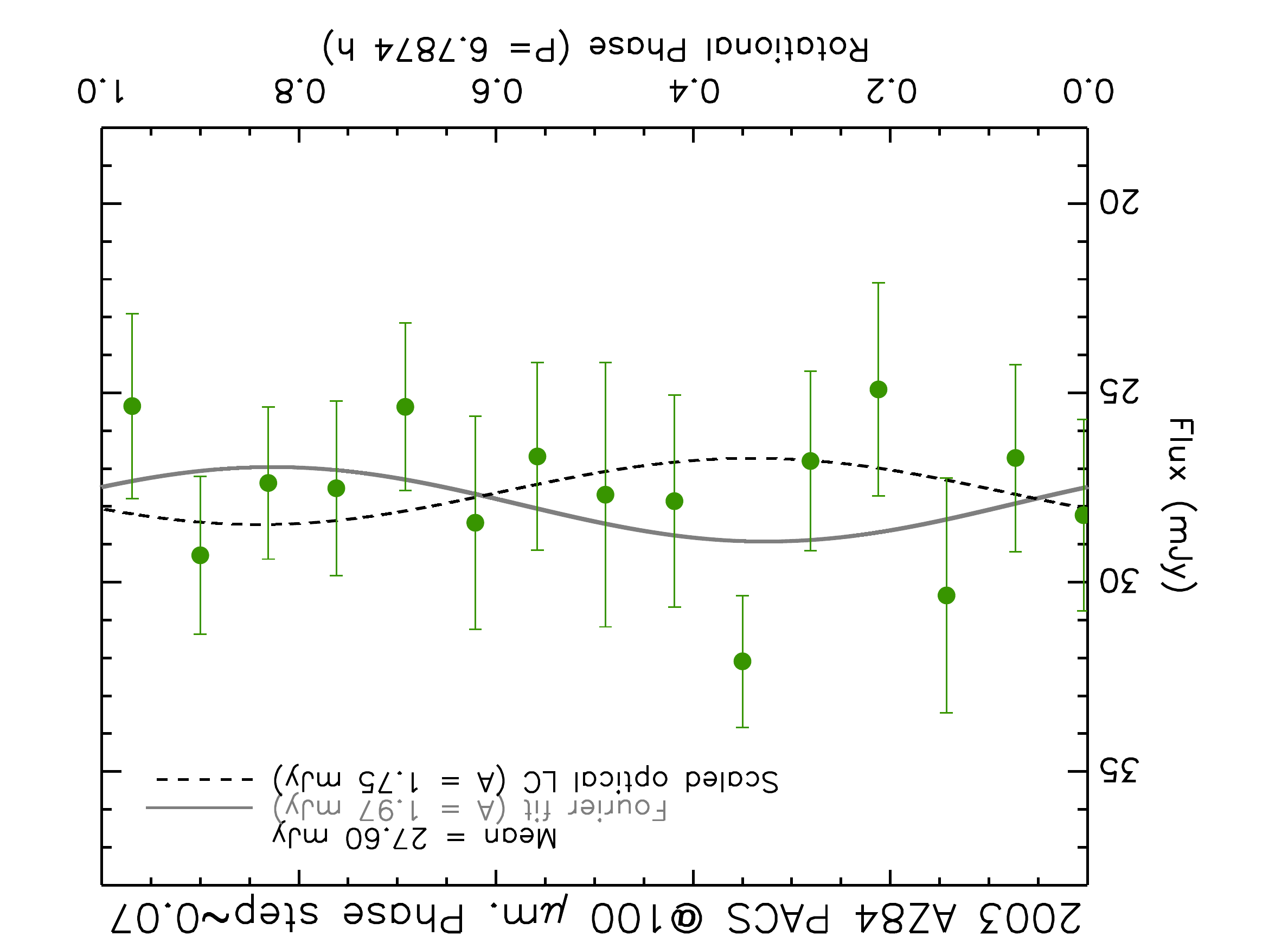}
  
      \caption{
Thermal time series curve at 100 $\mu$m for 2003 AZ$_{84}$. The black dashed curve is the scaled optical light curve obtained by rescaling the optical brightness to match the mean of the thermal fluxes. The grey solid curve is a Fourier fit to the thermal data. The reference for zero phase is at JD = 2453026.546400 days, uncorrected for light-time. Rotational phases have been computed using light-time corrected julian dates and light-time corrected zero date (see caption of table \ref{AZ84PACSdata} for further details). The uncertainty in the rotational phase is $\pm$ 0.01.
}

         \label{LC_az84green}
         
\end{figure}

The thermal light curve of 2003 AZ$_{84}$ is not firmly detected in the PACS data at 100 $\mu$m (see Fig. \ref{LC_az84green}). While a Fourier fit to the thermal data formally provides a best fit amplitude of 1.97 $\pm$ 1.40 mJy at 100 $\mu$m, its significance is thus at the 1.4 $\sigma$ level. By making a visual comparison of the Fourier fit to the thermal data in Fig. \ref{LC_az84green} with the Fourier fit to the optical data (shown with a dashed line), it appears that there could be a weak anticorrelation of the 100 $\mu$m data with the visible data. This would give confidence to the interpretation that the thermal light curve could be generated by enhanced thermal emission in the darker spots or darker terrains that give rise to the optical light curve, in the same way as the dark spot in Haumea generates enhanced thermal emission. However a Spearman test to analyze a possible anticorrelation of the thermal data with the optical light curve gave a non-significant result. Moreover, in the regime of low albedo ($\sim$ 10\% for 2003 AZ$_{84}$), the thermal emission is essentially albedo-independent, so that a thermal light curve resulting from optical markings would have an undetectable amplitude, barely above 0.05 mJy.

%______________________________________________________________

\subsubsection{Analysis of 2003 AZ$_{84}$ results} 

2003 AZ$_{84}$ is a large enough TNO \citep[D$_{equiv}$ = 727 km according][]{2012A&A...541A..93M} so that it is very likely to be in hydrostatic equilibrium. This means that the expected 3D shape of this TNO should be a figure of equilibrium: either a rotationally symmetric Maclaurin spheroid or a triaxial Jacobi body.  
 2003 AZ$_{84}$ has a low light curve amplitude in the visible, which means that either this TNO is a Maclaurin object with small albedo variability on its surface, or it is seen nearly pole on, or both. Recent results on stellar occultations by 2003 AZ$_{84}$ \citep{DiasOliveira2016} have shown an equal-area diameter of D$_{equiv}$ = 766 km and a small projected flattening of only 0.05. The small flattening has two possible extreme explanations: The object has a typical density of a TNO of its size but it is seen nearly pole-on so that the large flattening of the body becomes a small projected flattening, or 2003 AZ$_{84}$ has an exceptionally high density for its size. The density required for a Maclaurin body with flattening of 0.05 and a rotation period of 6.79 h is 5 500 kg/m$^{3}$. This huge density is not feasible in the Transneptunian region. Assuming that 2003 AZ$_{84}$ could have a density of $\sim$2500 kg/m$^{3}$, which is already too high for a TNO of its size, its true oblateness would be 0.12. This would require an aspect angle $<$ 45 degrees for the Maclaurin spheroid to give rise to the projected oblateness of 0.05 seen in the occultation. We note that densities of around 2 500 kg/m$^{3}$ have only been measured for the very largest TNOs, such as Pluto, Eris and Haumea whose internal pressures do not allow for the macroporosity that can exist in bodies of smaller size \citep[see e.g.][]{2002AJ....123.2110J}. So it is extremely unlikely that 2003 AZ$_{84}$ could have such a high density of 2 500 kg/m$^{3}$. Hence we are confident that the aspect angle of 2003 AZ$_{84}$ must be smaller than 45 degrees. Therefore the low light curve amplitude in the visible, the small thermal variability, and the occultation results are reasons to believe that 2003 AZ$_{84}$ could be close to pole-on (have a small aspect angle).

\begin{figure}[!hpbt]
   \centering
   
   \includegraphics[width=7.2cm,angle=90]{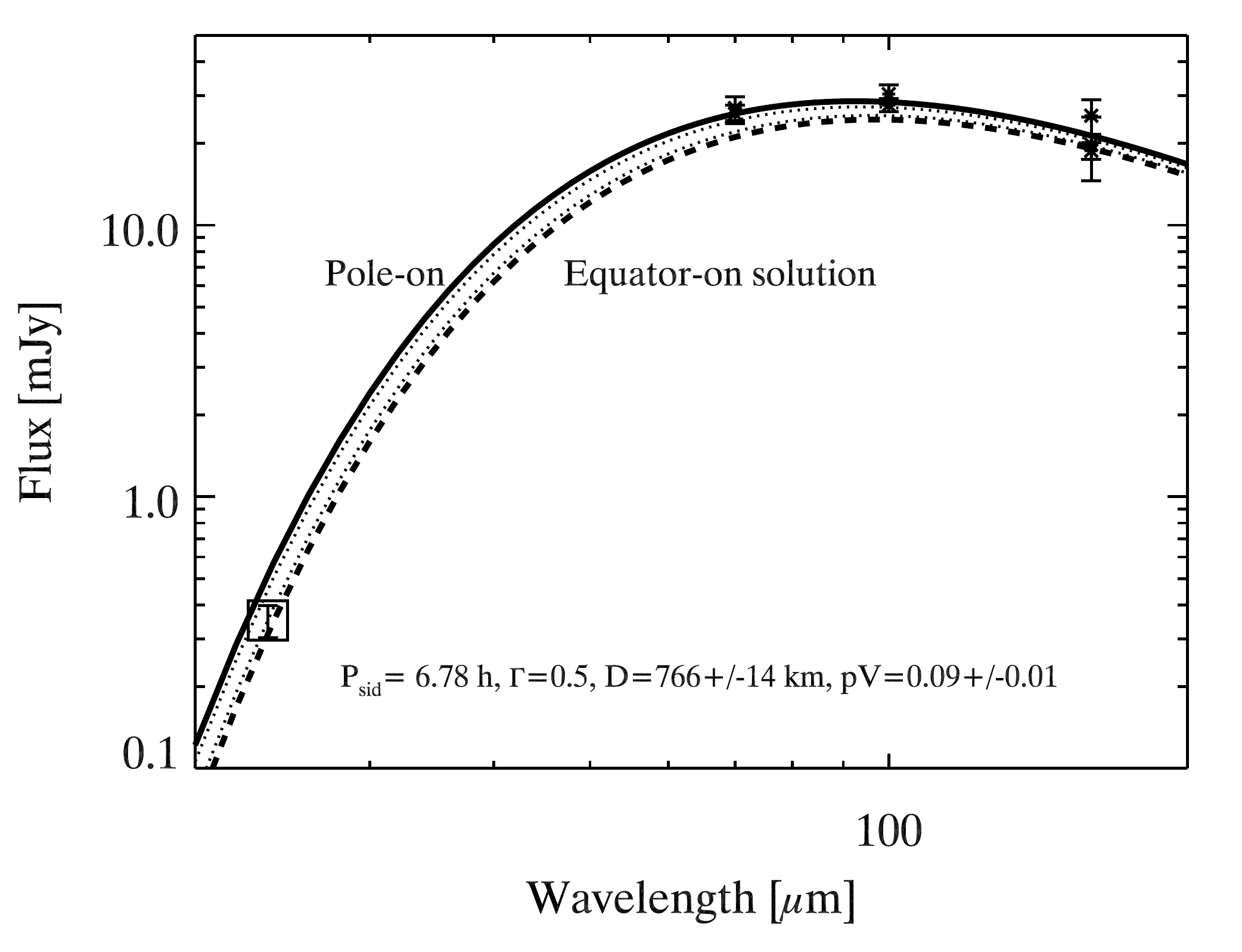}
 
      \caption{Absolute PACS and MIPS fluxes for 2003 AZ$_{84}$ with various thermophysical models overplotted: pole-on solution, pole-on +30$^{\circ}$, pole-on +60$^{\circ}$, equator-on. All the models use the occultation size and reasonable assumptions for the thermal properties.} 
      
         \label{AZ84_TPM_poleon}
         
\end{figure}

To further constrain the spin axis orientation, we run a thermophysical model \citep[TPM:][]{1996A&A...310.1011L,1997A&A...325.1226L,1998A&A...332.1123L}. The model takes into account the thermal conduction and surface roughness for objects of arbitrary shapes and spin properties. The model was extensively tested against thermal observations of near Earth asteroids (NEAs), Main Belt asteroids (MBAs), and TNOs over the last two decades. Recent works with this TPM code \citep[e.g.][]{2015A&A...583A..93P,2016AJ....151..117P,2017A&A...600A..12S,2013A&A...555A..15F} have shown that a combined analysis of Herschel and Spitzer thermal measurements enable us to constrain the spin-axis orientation of TNOs. Following up on this expertise, we also applied this code in the case of 2003 AZ$_{84}$, assuming a spherical shape model and a constant emissivity of 0.9 at all MIPS and PACS wavelengths \citep[emissivity effects are only expected at longer wavelengths, see][]{2013A&A...555A..15F}. We used all available Herschel/PACS and Spitzer/MIPS thermal data (except the Spitzer/MIPS data point at 71.42 $\mu$m, which is affected by a background source). The Spitzer MIPS data were presented in \cite{stansberry08}. We re-analysed the two MIPS observations of 2003 AZ$_{84}$. The 71.42 $\mu$m measurements are problematic owing to contaminating background sources, but the 23.68 $\mu$m points are clean and the object's point-spread-function is as expected. Table \ref{thermal_data_az84} shows the thermal measurements used in the thermophysical modeling. Using H$_{V}$ = 3.78 $\pm$ 0.05 mag and the stellar occultation size D = 766 $\pm$ 16 km \citep{DiasOliveira2016}, we check models with a range of thermal inertias from 0.0 to 100 MKS and a range of different levels of surface roughness (rms of surface slopes of 0.1, 0.3, 0.5, 0.7, and 0.9). The biggest issue with the radiometric analysis is that the MIPS and the PACS data do not match very well. A standard D-p$_{V}$ radiometric analysis for only the PACS data favors the pole-on solutions (combined with low surface roughness - rms of surface slopes $<$ 0.5) and provide the correct occultation size (760-790 km), for the two possible rotation periods (P = 6.7874 h and P = 13.5748 h) and almost independent of thermal inertia. The overall fit to only PACS observations is excellent. Equator-on solutions can also provide correct sizes, but only under the assumptions of extremely low thermal inertias far below 1.0 MKS and combined with extremely high surface roughness (rms of surface slopes $>$ 0.7), which looks very unrealistic. The very best solutions are found for a spin-axis orientation 30$^{\circ}$ away from pole-on, intermediate levels of surface roughness (i.e. realistics values), and acceptable values for the thermal inertia ($\Gamma$ = 0.5-3.0 MKS). Overall, the pole-on $\pm$30$^{\circ}$ configuration explains very well the PACS fluxes, but slightly overestimates the MIPS 24 $\mu$m within the 2 $\sigma$ level (see Fig. \ref{AZ84_TPM_poleon}). This difference between MIPS flux and model could be due to some light curve effects at the moment of the Spitzer/MIPS observations. The equator-on geometry only works when using very extreme settings, which seem very unrealistic. Summarizing the combined thermal and occultation analysis, we find that its spin-axis is very likely close to pole-on ($\pm$30$^{\circ}$).

\begin{table*}

\caption{Thermal data of 2003 AZ$_{84}$ used for the thermophysical modeling} % title of Table

\label{thermal_data_az84} % is used to refer this table in the text
 
\centering % used for centering table

\begin{tabular}{r c c c c} % centered columns (4 columns)

\hline\hline % inserts double horizontal lines

OBSID/AORKEY &  JD      & Band   &      Flux/unc & Telescope/instrument\\
        & [days] &   [$\mu$m]    &       [mJy]  &                                                               \\

% table heading

\hline % inserts single horizontal line
1342187054 & 2455152.31944 & 70.0  & 27.0 $\pm$ 2.7$^{(1)}$ & Herschel/PACS\\
1342187054 & 2455152.31944 & 160.0 & 19.8 $\pm$ 5.2$^{(1)}$ & Herschel/PACS\\
1342205152 & 2455466.80556 & 100.0 & 27.6 $\pm$ 1.5$^{(2)}$ & Herschel/PACS\\
1342205152 & 2455466.80556 & 160.0 & 18.8 $\pm$ 1.3$^{(2)}$ & Herschel/PACS\\
1342205223 & 2455467.62847 &  70.0 & 25.7 $\pm$ 2.0$^{(2)}$ & Herschel/PACS\\
1342205225 & 2455467.64375 & 100.0 & 30.4 $\pm$ 2.5$^{(2)}$ & Herschel/PACS\\
1342205225 & 2455467.64375 & 160.0 & 25.3 $\pm$ 3.7$^{(2)}$ & Herschel/PACS\\
10679040   & 2453824.88889 & 23.68 & 0.35  $\pm$ 0.047$^{(3)}$ & Spitzer/MIPS\\
                                                                
\hline %inserts single line

\end{tabular}

\footnotesize{OBSID is the Herschel internal observation ID. AORKEY is the Spitzer internal observation identification. JD is the Julian Date at the middle of integration uncorrected for light-time. Band is the PACS or MIPS bands in $\mu$m. Flux/unc are the color-corrected fluxes and uncertainties expressed in mJy. Telescope/instrument indicates the telescope (Herschel or Spitzer) and the instrument (PACS or MIPS). References: $^{(1)}$ chop/nod observations from \cite{2010A&A...518L.146M}; $^{(2)}$ This work; $^{(3)}$ Updated fluxes from \cite{stansberry08}.}

\end{table*}

%______________________________________________________________

\section{Summary and brief discussion}
\label{summary}

Time series thermal data of three bright TNOs (Haumea, 2003 VS$_2$ and 2003 AZ$_{84}$) have been acquired with Herschel/PACS in search of thermal light curves, with the  following main results:
  \begin{itemize}
   
                \item{The thermal light curve of Haumea is clearly detected at 100 and 160 $\mu$m, superseding the
early results of \cite{2010A&A...518L.147L} with Herschel/PACS. 
Both light curves are correlated with the optical one, implying primarily shape-driven light curves. Nonetheless,
the 100-$\mu$m data indicates a small extra flux at rotational phases affected by the optical dark spot.}

                \item{Thermophysical modeling of the Haumea thermal light curves indicates an overall surface 
with an extremely small thermal inertia ($\Gamma <$ 0.5 MKS and probably $\Gamma <$ 0.2 MKS) and high phase integral (q $\sim$ 0.8 for $\Gamma$ = 0.1 MKS and no surface roughness), which will be even higher if surface roughness is present.}

                \item{The energetic and thermophysical properties of Haumea's dark spot appear to be only modestly
different from the rest of the surfaces, with changes of only $\sim$ +0.05--0.1 MKS in thermal inertia or $\sim$ +0.1 in phase integral. We favor the case for a small increase of thermal inertia in the dark region.} 
                
                \item{The thermal light curve of 2003 VS$_{2}$ is not firmly detected at 70 $\mu$m and at 160 $\mu$m. However, Fourier fits to the thermal data are correlated with the optical light curve. The amplitude and mean flux of 2003 VS$_{2}$'s 70 $\mu$m light curve indicate a thermal inertia $\Gamma$ = (2.0$\pm$0.5) MKS.}
                
                \item{The thermal light curve of 2003 AZ$_{84}$ at 100 $\mu$m is not firmly detected. A thermophysical model applied to the mean thermal light curve fluxes and to all the Herschel/PACS and Spitzer/MIPS thermal data favors a close to pole-on ($\pm$30$^{\circ}$) orientation.}
            
      \item{Our conclusion of extremely small thermal inertias for 2003 VS$_2$ and even smaller for Haumea
statistically nicely matches  inferences on the TNO/Centaurs population based on Spitzer/Herschel
radiometry \citep{2013A&A...557A..60L}, including an albedo dependence of the thermal inertia.
These authors interpreted their results in terms of highly porous surfaces, in which the heat transfer 
efficiency is affected by radiative conductivity within pores and increases with depth in the subsurface. For heat conduction dominated by radiation, the thermal inertia is essentially proportional to $r_{h}^{-3/4}$ \citep{2015aste.book..107D}, or to $r_{h}^{-(0.9-1.0)}$ if the temperature dependence of the specific heat of ice is taken into account \citep{2013A&A...557A..60L}. Our thermal inertia for the three objects (2.0 $\pm$ 0.5 MKS for 2003 VS$_{2}$ at $r_{h}$ = 36.5 AU, $\sim$ 0.2 MKS for Haumea at $r_{h}$ = 51 AU, and 0.7-2.0 MKS for 2003 AZ$_{84}$ at $r_{h}$ = 45 AU) convert into $\Gamma$ = 10--90 MKS and $\Gamma$ = 4--35 MKS at 1 AU for the two temperature-dependence cases, respectively. While somewhat even lower, these numbers compare generally well with the thermal inertias of large ($>$ 100 km) asteroids \citep[10-300 MKS, corrected to 1 AU, see Fig. 9 in][]{2015aste.book..107D}, where the smallest values are indicative of fine grain regolith. Recently, \cite{2016A&A...588A.133F} re-addressed the general issue of low thermal inertias
in outer solar system bodies (including icy satellites), and pointed out several other
important factors, in addition to surface porosity, affecting surface effective thermal inertias.
One such factor is the quality of grain contact (i.e. tight or loose) in determining solid-state conductivity.
For H$_2$O-ice covered surfaces, another factor, already recognized by \cite{2013A&A...557A..60L}, is
the amorphous vs. crystalline state of water, as the two states are associated with different bulk
conductivities (and different temperature dependence thereof). On the basis of a detailed physical model
of conductivity, including radiative conductivity, \cite{2016A&A...588A.133F}
were able to reproduce the order of magnitude and heliocentric dependence of the thermal inertias measured
by \cite{2013A&A...557A..60L}, by invoking loose contacts in a moderately porous regolith of sub-cm-sized grains
made of amorphous ice. Since water ice, when detected on the surface of TNOs, is usually in cristalline form,
this scenario implies the presence of amorphous ice at cm depths below a thin layer of crystalline ice.}

  \end{itemize}

%______________________________________________________________

\begin{acknowledgements}

\textit{Herschel} is an ESA space observatory with science instruments provided by European led Principal Investigator consortia and with important participation from NASA. \textit{Herschel} data presented in this work were processed using HIPE, a joint development by the \textit{Herschel Science Ground Segment Consortium}, consisting of ESA, the NASA \textit{Herschel Science Center}, and the HIFI, PACS and SPIRE consortia. This research is partially based on observations collected at Centro Astron{\'o}mico Hispano Alem{\'a}n (CAHA) at Calar Alto, operated jointly by the Max-Planck Institut fur Astronomie and the Instituto de Astrof\'{\i}sica de Andaluc\'{\i}a (CSIC). This research is partially based as well on observations carried out at the Observatorio de Sierra Nevada (OSN) operated by Instituto de Astrof\'{\i}sica de Andaluc\'{\i}a (CSIC). The research leading to these results has received funding from the European Union's Horizon 2020 Research and Innovation Programme, under Grant Agreement no 687378. P. Santos-Sanz and J.L. Ortiz would like to acknowledge financial support by the Spanish grant AYA-2014-56637-C2-1-P and the Proyecto de Excelencia de la Junta de Andaluc\'{\i}a J.A. 2012-FQM1776. C. Kiss acknowledges financial support from NKFIH grant GINOP-2.3.2-15-2016-00003. E. Vilenius was supported by the German DLR project number 50 OR 1108. R. Duffard acknowledges financial support from the MINECO for his Ramon y Cajal Contract. 
     
\end{acknowledgements}

% WARNING
%-------------------------------------------------------------------
% Please note that we have included the references to the file aa.dem in
% order to compile it, but we ask you to:
%
% - use BibTeX with the regular commands:
%   \bibliographystyle{aa} % style aa.bst
%   \bibliography{Yourfile} % your references Yourfile.bib
%
% - join the .bib files when you upload your source files
%-------------------------------------------------------------------

\bibliographystyle{aa}
%\bibliography{thermal_lcs}

\begin{thebibliography}{68}
\expandafter\ifx\csname natexlab\endcsname\relax\def\natexlab#1{#1}\fi

\bibitem[{{Barkume} {et~al.}(2008){Barkume}, {Brown}, \&
  {Schaller}}]{2008AJ....135...55B}
{Barkume}, K.~M., {Brown}, M.~E., \& {Schaller}, E.~L. 2008, \aj, 135, 55

\bibitem[{{Barucci} {et~al.}(2011){Barucci}, {Alvarez-Candal}, {Merlin},
  {Belskaya}, {de Bergh}, {Perna}, {DeMeo}, \&
  {Fornasier}}]{2011Icar..214..297B}
{Barucci}, M.~A., {Alvarez-Candal}, A., {Merlin}, F., {et~al.} 2011, \icarus,
  214, 297

\bibitem[{{Brown} {et~al.}(2007){Brown}, {Barkume}, {Ragozzine}, \&
  {Schaller}}]{2007Natur.446..294B}
{Brown}, M.~E., {Barkume}, K.~M., {Ragozzine}, D., \& {Schaller}, E.~L. 2007,
  \nat, 446, 294

\bibitem[{{Brucker} {et~al.}(2009){Brucker}, {Grundy}, {Stansberry}, {Spencer},
  {Sheppard}, {Chiang}, \& {Buie}}]{Brucker2009}
{Brucker}, M.~J., {Grundy}, W.~M., {Stansberry}, J.~A., {et~al.} 2009, \icarus,
  201, 284

\bibitem[{{Chandrasekhar}(1987)}]{Chandra87}
{Chandrasekhar}, S. 1987, {Ellipsoidal figures of equilibrium}

\bibitem[{{Delbo} {et~al.}(2015){Delbo}, {Mueller}, {Emery}, {Rozitis}, \&
  {Capria}}]{2015aste.book..107D}
{Delbo}, M., {Mueller}, M., {Emery}, J.~P., {Rozitis}, B., \& {Capria}, M.~T.
  2015, {Asteroid Thermophysical Modeling}, ed. P.~{Michel}, F.~E. {DeMeo}, \&
  W.~F. {Bottke}, 107--128

\bibitem[{{Dias-Oliveira et al.}(2016)}]{DiasOliveira2016}
{Dias-Oliveira et al.} 2016, submitted to \apj

\bibitem[{{Duffard} {et~al.}(2009){Duffard}, {Ortiz}, {Thirouin},
  {Santos-Sanz}, \& {Morales}}]{2009A&A...505.1283D}
{Duffard}, R., {Ortiz}, J.~L., {Thirouin}, A., {Santos-Sanz}, P., \& {Morales},
  N. 2009, \aap, 505, 1283

\bibitem[{{Duffard} {et~al.}(2014{\natexlab{a}}){Duffard}, {Pinilla-Alonso},
  {Ortiz}, {Alvarez-Candal}, {Sicardy}, {Santos-Sanz}, {Morales}, {Colazo},
  {Fern{\'a}ndez-Valenzuela}, \& {Braga-Ribas}}]{2014A&A...568A..79D}
{Duffard}, R., {Pinilla-Alonso}, N., {Ortiz}, J.~L., {et~al.}
  2014{\natexlab{a}}, \aap, 568, A79

\bibitem[{{Duffard} {et~al.}(2014{\natexlab{b}}){Duffard}, {Pinilla-Alonso},
  {Santos-Sanz}, {Vilenius}, {Ortiz}, {Mueller}, {Fornasier}, {Lellouch},
  {Mommert}, {Pal}, {Kiss}, {Mueller}, {Stansberry}, {Delsanti}, {Peixinho}, \&
  {Trilling}}]{2014A&A...564A..92D}
{Duffard}, R., {Pinilla-Alonso}, N., {Santos-Sanz}, P., {et~al.}
  2014{\natexlab{b}}, \aap, 564, A92

\bibitem[{{Fernandez-Valenzuela} {et~al.}(2017){Fernandez-Valenzuela}, {Ortiz},
  {Duffard}, {Morales}, \& {Santos-Sanz}}]{estela2016}
{Fernandez-Valenzuela}, E., {Ortiz}, J.~L., {Duffard}, R., {Morales}, N., \&
  {Santos-Sanz}, P. 2017, MNRAS, 466, 4147

\bibitem[{{Ferrari} \& {Lucas}(2016)}]{2016A&A...588A.133F}
{Ferrari}, C. \& {Lucas}, A. 2016, \aap, 588, A133

\bibitem[{{Fornasier} {et~al.}(2013){Fornasier}, {Lellouch}, {M{\"u}ller},
  {Santos-Sanz}, {Panuzzo}, {Kiss}, {Lim}, {Mommert}, {Bockel{\'e}e-Morvan},
  {Vilenius}, {Stansberry}, {Tozzi}, {Mottola}, {Delsanti}, {Crovisier},
  {Duffard}, {Henry}, {Lacerda}, {Barucci}, \& {Gicquel}}]{2013A&A...555A..15F}
{Fornasier}, S., {Lellouch}, E., {M{\"u}ller}, T., {et~al.} 2013, \aap, 555,
  A15

\bibitem[{{Gladman} {et~al.}(2008){Gladman}, {Marsden}, \&
  {Vanlaerhoven}}]{2008ssbn.book...43G}
{Gladman}, B., {Marsden}, B.~G., \& {Vanlaerhoven}, C. 2008, {Nomenclature in
  the Outer Solar System}, ed. M.~A. {Barucci}, H.~{Boehnhardt}, D.~P.
  {Cruikshank}, A.~{Morbidelli}, \& R.~{Dotson}, 43--57

\bibitem[{{Griffin} {et~al.}(2010){Griffin}, {Abergel}, {Abreu}, {Ade},
  {Andr{\'e}}, {Augueres}, {Babbedge}, {Bae}, {Baillie}, {Baluteau}, {Barlow},
  {Bendo}, {Benielli}, {Bock}, {Bonhomme}, {Brisbin}, {Brockley-Blatt},
  {Caldwell}, {Cara}, {Castro-Rodriguez}, {Cerulli}, {Chanial}, {Chen},
  {Clark}, {Clements}, {Clerc}, {Coker}, {Communal}, {Conversi}, {Cox},
  {Crumb}, {Cunningham}, {Daly}, {Davis}, {de Antoni}, {Delderfield}, {Devin},
  {di Giorgio}, {Didschuns}, {Dohlen}, {Donati}, {Dowell}, {Dowell}, {Duband},
  {Dumaye}, {Emery}, {Ferlet}, {Ferrand}, {Fontignie}, {Fox}, {Franceschini},
  {Frerking}, {Fulton}, {Garcia}, {Gastaud}, {Gear}, {Glenn}, {Goizel},
  {Griffin}, {Grundy}, {Guest}, {Guillemet}, {Hargrave}, {Harwit}, {Hastings},
  {Hatziminaoglou}, {Herman}, {Hinde}, {Hristov}, {Huang}, {Imhof}, {Isaak},
  {Israelsson}, {Ivison}, {Jennings}, {Kiernan}, {King}, {Lange}, {Latter},
  {Laurent}, {Laurent}, {Leeks}, {Lellouch}, {Levenson}, {Li}, {Li},
  {Lilienthal}, {Lim}, {Liu}, {Lu}, {Madden}, {Mainetti}, {Marliani}, {McKay},
  {Mercier}, {Molinari}, {Morris}, {Moseley}, {Mulder}, {Mur}, {Naylor},
  {Nguyen}, {O'Halloran}, {Oliver}, {Olofsson}, {Olofsson}, {Orfei}, {Page},
  {Pain}, {Panuzzo}, {Papageorgiou}, {Parks}, {Parr-Burman}, {Pearce},
  {Pearson}, {P{\'e}rez-Fournon}, {Pinsard}, {Pisano}, {Podosek}, {Pohlen},
  {Polehampton}, {Pouliquen}, {Rigopoulou}, {Rizzo}, {Roseboom}, {Roussel},
  {Rowan-Robinson}, {Rownd}, {Saraceno}, {Sauvage}, {Savage}, {Savini},
  {Sawyer}, {Scharmberg}, {Schmitt}, {Schneider}, {Schulz}, {Schwartz},
  {Shafer}, {Shupe}, {Sibthorpe}, {Sidher}, {Smith}, {Smith}, {Smith},
  {Spencer}, {Stobie}, {Sudiwala}, {Sukhatme}, {Surace}, {Stevens}, {Swinyard},
  {Trichas}, {Tourette}, {Triou}, {Tseng}, {Tucker}, {Turner}, {Vaccari},
  {Valtchanov}, {Vigroux}, {Virique}, {Voellmer}, {Walker}, {Ward}, {Waskett},
  {Weilert}, {Wesson}, {White}, {Whitehouse}, {Wilson}, {Winter}, {Woodcraft},
  {Wright}, {Xu}, {Zavagno}, {Zemcov}, {Zhang}, \&
  {Zonca}}]{2010A&A...518L...3G}
{Griffin}, M.~J., {Abergel}, A., {Abreu}, A., {et~al.} 2010, \aap, 518, L3

\bibitem[{{Groussin} {et~al.}(2004){Groussin}, {Lamy}, \&
  {Jorda}}]{2004A&A...413.1163G}
{Groussin}, O., {Lamy}, P., \& {Jorda}, L. 2004, \aap, 413, 1163

\bibitem[{{Guilbert} {et~al.}(2009){Guilbert}, {Alvarez-Candal}, {Merlin},
  {Barucci}, {Dumas}, {de Bergh}, \& {Delsanti}}]{2009Icar..201..272G}
{Guilbert}, A., {Alvarez-Candal}, A., {Merlin}, F., {et~al.} 2009, \icarus,
  201, 272

\bibitem[{{Harris}(1998)}]{Harris98}
{Harris}, A.~W. 1998, \icarus, 131, 291

\bibitem[{{Hillier} {et~al.}(1991){Hillier}, {Helfenstein}, {Verbiscer}, \&
  {Veverka}}]{1991JGR....9619203H}
{Hillier}, J., {Helfenstein}, P., {Verbiscer}, A., \& {Veverka}, J. 1991, \jgr,
  96, 19

\bibitem[{{Howell}(1989)}]{1989PASP..101..616H}
{Howell}, S.~B. 1989, \pasp, 101, 616

\bibitem[{{Jewitt} {et~al.}(2007){Jewitt}, {Peixinho}, \&
  {Hsieh}}]{2007AJ....134.2046J}
{Jewitt}, D., {Peixinho}, N., \& {Hsieh}, H.~H. 2007, \aj, 134, 2046

\bibitem[{{Jewitt} \& {Sheppard}(2002)}]{2002AJ....123.2110J}
{Jewitt}, D.~C. \& {Sheppard}, S.~S. 2002, \aj, 123, 2110

\bibitem[{{Jorda} {et~al.}(2010){Jorda}, {Spjuth}, {Keller}, {Lamy}, \&
  {Llebaria}}]{2010SPIE.7533E..11J}
{Jorda}, L., {Spjuth}, S., {Keller}, H.~U., {Lamy}, P., \& {Llebaria}, A. 2010,
  in \procspie, Vol. 7533, Computational Imaging VIII, 753311

\bibitem[{{Kiss} {et~al.}(2014){Kiss}, {M{\"u}ller}, {Vilenius}, {P{\'a}l},
  {Santos-Sanz}, {Lellouch}, {Marton}, {Vereb{\'e}lyi}, {Szalai}, {Hartogh},
  {Stansberry}, {Henry}, \& {Delsanti}}]{2013ExA...tmp...36K}
{Kiss}, C., {M{\"u}ller}, T.~G., {Vilenius}, E., {et~al.} 2014, Experimental
  Astronomy, 37, 161

\bibitem[{{Kiss et al.}(2016)}]{kiss2016}
{Kiss et al.} 2016, submitted to \aap

\bibitem[{{Lacerda} {et~al.}(2008){Lacerda}, {Jewitt}, \&
  {Peixinho}}]{2008AJ....135.1749L}
{Lacerda}, P., {Jewitt}, D., \& {Peixinho}, N. 2008, \aj, 135, 1749

\bibitem[{{Lacerda} \& {Jewitt}(2007)}]{2007AJ....133.1393L}
{Lacerda}, P. \& {Jewitt}, D.~C. 2007, \aj, 133, 1393

\bibitem[{{Lagerros}(1996)}]{1996A&A...310.1011L}
{Lagerros}, J.~S.~V. 1996, \aap, 310, 1011

\bibitem[{{Lagerros}(1997)}]{1997A&A...325.1226L}
{Lagerros}, J.~S.~V. 1997, \aap, 325, 1226

\bibitem[{{Lagerros}(1998)}]{1998A&A...332.1123L}
{Lagerros}, J.~S.~V. 1998, \aap, 332, 1123

\bibitem[{{Lellouch} {et~al.}(2010){Lellouch}, {Kiss}, {Santos-Sanz},
  {M{\"u}ller}, {Fornasier}, {Groussin}, {Lacerda}, {Ortiz}, {Thirouin},
  {Delsanti}, {Duffard}, {Harris}, {Henry}, {Lim}, {Moreno}, {Mommert},
  {Mueller}, {Protopapa}, {Stansberry}, {Trilling}, {Vilenius}, {Barucci},
  {Crovisier}, {Doressoundiram}, {Dotto}, {Guti{\'e}rrez}, {Hainaut},
  {Hartogh}, {Hestroffer}, {Horner}, {Jorda}, {Kidger}, {Lara}, {Rengel},
  {Swinyard}, \& {Thomas}}]{2010A&A...518L.147L}
{Lellouch}, E., {Kiss}, C., {Santos-Sanz}, P., {et~al.} 2010, \aap, 518, L147

\bibitem[{{Lellouch} {et~al.}(2000){Lellouch}, {Laureijs}, {Schmitt},
  {Quirico}, {de Bergh}, {Crovisier}, \& {Coustenis}}]{2000Icar..147..220L}
{Lellouch}, E., {Laureijs}, R., {Schmitt}, B., {et~al.} 2000, \icarus, 147, 220

\bibitem[{{Lellouch} {et~al.}(2016){Lellouch}, {Santos-Sanz}, {Fornasier},
  {Lim}, {Stansberry}, {Vilenius}, {Kiss}, {M{\"u}ller}, {Marton}, {Protopapa},
  {Panuzzo}, \& {Moreno}}]{LellouchPlutoLCs}
{Lellouch}, E., {Santos-Sanz}, P., {Fornasier}, S., {et~al.} 2016, \aap, 588,
  A2

\bibitem[{{Lellouch} {et~al.}(2013){Lellouch}, {Santos-Sanz}, {Lacerda},
  {Mommert}, {Duffard}, {Ortiz}, {M{\"u}ller}, {Fornasier}, {Stansberry},
  {Kiss}, {Vilenius}, {Mueller}, {Peixinho}, {Moreno}, {Groussin}, {Delsanti},
  \& {Harris}}]{2013A&A...557A..60L}
{Lellouch}, E., {Santos-Sanz}, P., {Lacerda}, P., {et~al.} 2013, \aap, 557, A60

\bibitem[{{Lellouch} {et~al.}(2011){Lellouch}, {Stansberry}, {Emery}, {Grundy},
  \& {Cruikshank}}]{2011Icar..214..701L}
{Lellouch}, E., {Stansberry}, J., {Emery}, J., {Grundy}, W., \& {Cruikshank},
  D.~P. 2011, \icarus, 214, 701

\bibitem[{{Lim} {et~al.}(2010){Lim}, {Stansberry}, {M{\"u}ller}, {Mueller},
  {Lellouch}, {Kiss}, {Santos-Sanz}, {Vilenius}, {Protopapa}, {Moreno},
  {Delsanti}, {Duffard}, {Fornasier}, {Groussin}, {Harris}, {Henry}, {Horner},
  {Lacerda}, {Mommert}, {Ortiz}, {Rengel}, {Thirouin}, {Trilling}, {Barucci},
  {Crovisier}, {Doressoundiram}, {Dotto}, {Guti{\'e}rrez Buenestado},
  {Hainaut}, {Hartogh}, {Hestroffer}, {Kidger}, {Lara}, {Swinyard}, \&
  {Thomas}}]{2010A&A...518L.148L}
{Lim}, T.~L., {Stansberry}, J., {M{\"u}ller}, T.~G., {et~al.} 2010, \aap, 518,
  L148

\bibitem[{{Lockwood} \& {Brown}(2009)}]{2009DPS....41.6506L}
{Lockwood}, A. \& {Brown}, M.~E. 2009, in AAS/Division for Planetary Sciences
  Meeting Abstracts, Vol.~41, AAS/Division for Planetary Sciences Meeting
  Abstracts \#41, 65.06

\bibitem[{{Lockwood} {et~al.}(2014){Lockwood}, {Brown}, \&
  {Stansberry}}]{2014EM&P..tmp....3L}
{Lockwood}, A.~C., {Brown}, M.~E., \& {Stansberry}, J. 2014, Earth Moon and
  Planets, 111, 127

\bibitem[{{Mommert} {et~al.}(2012){Mommert}, {Harris}, {Kiss}, {P{\'a}l},
  {Santos-Sanz}, {Stansberry}, {Delsanti}, {Vilenius}, {M{\"u}ller},
  {Peixinho}, {Lellouch}, {Szalai}, {Henry}, {Duffard}, {Fornasier}, {Hartogh},
  {Mueller}, {Ortiz}, {Protopapa}, {Rengel}, \&
  {Thirouin}}]{2012A&A...541A..93M}
{Mommert}, M., {Harris}, A.~W., {Kiss}, C., {et~al.} 2012, \aap, 541, A93

\bibitem[{{M{\"u}ller} {et~al.}(2009){M{\"u}ller}, {Lellouch}, {B{\"o}hnhardt},
  {Stansberry}, {Barucci}, {Crovisier}, {Delsanti}, {Doressoundiram}, {Dotto},
  {Duffard}, {Fornasier}, {Groussin}, {Guti{\'e}rrez}, {Hainaut}, {Harris},
  {Hartogh}, {Hestroffer}, {Horner}, {Jewitt}, {Kidger}, {Kiss}, {Lacerda},
  {Lara}, {Lim}, {Mueller}, {Moreno}, {Ortiz}, {Rengel}, {Santos-Sanz},
  {Swinyard}, {Thomas}, {Thirouin}, \& {Trilling}}]{2009EM&P..105..209M}
{M{\"u}ller}, T.~G., {Lellouch}, E., {B{\"o}hnhardt}, H., {et~al.} 2009, Earth
  Moon and Planets, 105, 209

\bibitem[{{M{\"u}ller} {et~al.}(2010){M{\"u}ller}, {Lellouch}, {Stansberry},
  {Kiss}, {Santos-Sanz}, {Vilenius}, {Protopapa}, {Moreno}, {Mueller},
  {Delsanti}, {Duffard}, {Fornasier}, {Groussin}, {Harris}, {Henry}, {Horner},
  {Lacerda}, {Lim}, {Mommert}, {Ortiz}, {Rengel}, {Thirouin}, {Trilling},
  {Barucci}, {Crovisier}, {Doressoundiram}, {Dotto}, {Guti{\'e}rrez},
  {Hainaut}, {Hartogh}, {Hestroffer}, {Kidger}, {Lara}, {Swinyard}, \&
  {Thomas}}]{2010A&A...518L.146M}
{M{\"u}ller}, T.~G., {Lellouch}, E., {Stansberry}, J., {et~al.} 2010, \aap,
  518, L146

\bibitem[{{Ortiz} {et~al.}(2015){Ortiz}, {Duffard}, {Pinilla-Alonso},
  {Alvarez-Candal}, {Santos-Sanz}, {Morales}, {Fern{\'a}ndez-Valenzuela},
  {Licandro}, {Campo Bagatin}, \& {Thirouin}}]{Ortiz2015}
{Ortiz}, J.~L., {Duffard}, R., {Pinilla-Alonso}, N., {et~al.} 2015, \aap, 576,
  A18

\bibitem[{{Ortiz} {et~al.}(2006){Ortiz}, {Guti{\'e}rrez}, {Santos-Sanz},
  {Casanova}, \& {Sota}}]{2006A&A...447.1131O}
{Ortiz}, J.~L., {Guti{\'e}rrez}, P.~J., {Santos-Sanz}, P., {Casanova}, V., \&
  {Sota}, A. 2006, \aap, 447, 1131

\bibitem[{{Ortiz} {et~al.}(2012){Ortiz}, {Thirouin}, {Campo Bagatin},
  {Duffard}, {Licandro}, {Richardson}, {Santos-Sanz}, {Morales}, \&
  {Benavidez}}]{2012MNRAS.419.2315O}
{Ortiz}, J.~L., {Thirouin}, A., {Campo Bagatin}, A., {et~al.} 2012, \mnras,
  419, 2315
  
\bibitem[{{P{\'a}l} {et~al.}(2012){P{\'a}l}, {Kiss}, {M{\"u}ller},
  {Santos-Sanz}, {Vilenius}, {Szalai}, {Mommert}, {Lellouch}, {Rengel},
  {Hartogh}, {Protopapa}, {Stansberry}, {Ortiz}, {Duffard}, {Thirouin},
  {Henry}, \& {Delsanti}}]{2012A&A...541L...6P}
{P{\'a}l}, A., {Kiss}, C., {M{\"u}ller}, T.~G., {et~al.} 2012, \aap, 541, L6

\bibitem[{{P{\'a}l} {et~al.}(2015){P{\'a}l}, {Kiss}, {Horner}, {Szak{\'a}ts},
  {Vilenius}, {M{\"u}ller}, {Acosta-Pulido}, {Licandro}, {Cabrera-Lavers},
  {S{\'a}rneczky}, {Szab{\'o}}, {Thirouin}, {Sip{\H o}cz}, {D{\'o}zsa}, \&
  {Duffard}}]{2015A&A...583A..93P}
{P{\'a}l}, A., {Kiss}, C., {Horner}, J., {et~al.} 2015, \aap, 583, A93

\bibitem[{{P{\'a}l} {et~al.}(2016){P{\'a}l}, {Kiss}, {M{\"u}ller},
  {Moln{\'a}r}, {Szab{\'o}}, {Szab{\'o}}, {S{\'a}rneczky}, \&
  {Kiss}}]{2016AJ....151..117P}
{P{\'a}l}, A., {Kiss}, C., {M{\"u}ller}, T.~G., {et~al.} 2016, \aj, 151, 117

\bibitem[{{Perna} {et~al.}(2010){Perna}, {Barucci}, {Fornasier}, {DeMeo},
  {Alvarez-Candal}, {Merlin}, {Dotto}, {Doressoundiram}, \& {de
  Bergh}}]{2010A&A...510A..53P}
{Perna}, D., {Barucci}, M.~A., {Fornasier}, S., {et~al.} 2010, \aap, 510, A53

\bibitem[{{Pilbratt} {et~al.}(2010){Pilbratt}, {Riedinger}, {Passvogel},
  {Crone}, {Doyle}, {Gageur}, {Heras}, {Jewell}, {Metcalfe}, {Ott}, \&
  {Schmidt}}]{2010A&A...518L...1P}
{Pilbratt}, G.~L., {Riedinger}, J.~R., {Passvogel}, T., {et~al.} 2010, \aap,
  518, L1

\bibitem[{{Poglitsch} {et~al.}(2010){Poglitsch}, {Waelkens}, {Geis},
  {Feuchtgruber}, {Vandenbussche}, {Rodriguez}, {Krause}, {Renotte}, {van
  Hoof}, {Saraceno}, {Cepa}, {Kerschbaum}, {Agn{\`e}se}, {Ali}, {Altieri},
  {Andreani}, {Augueres}, {Balog}, {Barl}, {Bauer}, {Belbachir}, {Benedettini},
  {Billot}, {Boulade}, {Bischof}, {Blommaert}, {Callut}, {Cara}, {Cerulli},
  {Cesarsky}, {Contursi}, {Creten}, {De Meester}, {Doublier}, {Doumayrou},
  {Duband}, {Exter}, {Genzel}, {Gillis}, {Gr{\"o}zinger}, {Henning},
  {Herreros}, {Huygen}, {Inguscio}, {Jakob}, {Jamar}, {Jean}, {de Jong},
  {Katterloher}, {Kiss}, {Klaas}, {Lemke}, {Lutz}, {Madden}, {Marquet},
  {Martignac}, {Mazy}, {Merken}, {Montfort}, {Morbidelli}, {M{\"u}ller},
  {Nielbock}, {Okumura}, {Orfei}, {Ottensamer}, {Pezzuto}, {Popesso},
  {Putzeys}, {Regibo}, {Reveret}, {Royer}, {Sauvage}, {Schreiber}, {Stegmaier},
  {Schmitt}, {Schubert}, {Sturm}, {Thiel}, {Tofani}, {Vavrek}, {Wetzstein},
  {Wieprecht}, \& {Wiezorrek}}]{2010A&A...518L...2P}
{Poglitsch}, A., {Waelkens}, C., {Geis}, N., {et~al.} 2010, \aap, 518, L2

\bibitem[{{Rabinowitz} {et~al.}(2006){Rabinowitz}, {Barkume}, {Brown}, {Roe},
  {Schwartz}, {Tourtellotte}, \& {Trujillo}}]{2006ApJ...639.1238R}
{Rabinowitz}, D.~L., {Barkume}, K., {Brown}, M.~E., {et~al.} 2006, \apj, 639,
  1238

\bibitem[{{Rabinowitz} {et~al.}(2007){Rabinowitz}, {Schaefer}, \&
  {Tourtellotte}}]{2007AJ....133...26R}
{Rabinowitz}, D.~L., {Schaefer}, B.~E., \& {Tourtellotte}, S.~W. 2007, \aj,
  133, 26

\bibitem[{{Ragozzine} \& {Brown}(2009)}]{2009AJ....137.4766R}
{Ragozzine}, D. \& {Brown}, M.~E. 2009, \aj, 137, 4766

\bibitem[{{Santos-Sanz} {et~al.}(2012){Santos-Sanz}, {Lellouch}, {Fornasier},
  {Kiss}, {Pal}, {M{\"u}ller}, {Vilenius}, {Stansberry}, {Mommert}, {Delsanti},
  {Mueller}, {Peixinho}, {Henry}, {Ortiz}, {Thirouin}, {Protopapa}, {Duffard},
  {Szalai}, {Lim}, {Ejeta}, {Hartogh}, {Harris}, \&
  {Rengel}}]{2012A&A...541A..92S}
{Santos-Sanz}, P., {Lellouch}, E., {Fornasier}, S., {et~al.} 2012, \aap, 541,
  A92

\bibitem[{{Santos-Sanz} {et~al.}(2014){Santos-Sanz}, {Lellouch}, {Ortiz},
  {Kiss}, {M{\"u}ller}, {Vilenius}, {Stansberry}, {Fornasier}, {Lim},
  {Duffard}, {Lacerda}, \& {Thirouin}}]{SantosSanz2014}
{Santos-Sanz}, P., {Lellouch}, E., {Ortiz}, J.~L., {et~al.} 2014, European
  Planetary Science Congress 2014, EPSC Abstracts, Vol.~9, id.~EPSC2014-187, 9,
  EPSC2014

\bibitem[{{Schindler} {et~al.}(2017){Schindler}, {Wolf}, {Bardecker}, {Olsen},
  {M{\"u}ller}, {Kiss}, {Ortiz}, {Braga-Ribas}, {Camargo}, {Herald}, \&
  {Krabbe}}]{2017A&A...600A..12S}
{Schindler}, K., {Wolf}, J., {Bardecker}, J., {et~al.} 2017, \aap, 600, A12

\bibitem[{{Sheppard}(2007)}]{2007AJ....134..787S}
{Sheppard}, S.~S. 2007, \aj, 134, 787

\bibitem[{{Sheppard} \& {Jewitt}(2003)}]{2003EM&P...92..207S}
{Sheppard}, S.~S. \& {Jewitt}, D.~C. 2003, Earth Moon and Planets, 92, 207

\bibitem[{{Sheppard} {et~al.}(2008){Sheppard}, {Lacerda}, \&
  {Ortiz}}]{2008ssbn.book..129S}
{Sheppard}, S.~S., {Lacerda}, P., \& {Ortiz}, J.~L. 2008, {Photometric
  Lightcurves of Transneptunian Objects and Centaurs: Rotations, Shapes, and
  Densities}, ed. M.~A. {Barucci}, H.~{Boehnhardt}, D.~P. {Cruikshank},
  A.~{Morbidelli}, \& R.~{Dotson}, 129--142

\bibitem[{{Stansberry} {et~al.}(2008){Stansberry}, {Grundy}, {Brown},
  {Cruikshank}, {Spencer}, {Trilling}, \& {Margot}}]{stansberry08}
{Stansberry}, J., {Grundy}, W., {Brown}, M., {et~al.} 2008, {Physical
  Properties of Kuiper Belt and Centaur Objects: Constraints from the Spitzer
  Space Telescope}, ed. M.~A. {Barucci}, H.~{Boehnhardt}, D.~P. {Cruikshank},
  A.~{Morbidelli}, \& R.~{Dotson}, 161--179

\bibitem[{{Stetson}(1987)}]{1987PASP...99..191S}
{Stetson}, P.~B. 1987, \pasp, 99, 191

\bibitem[{{Tegler} {et~al.}(2005){Tegler}, {Romanishin}, {Consolmagno}, {Rall},
  {Worhatch}, {Nelson}, \& {Weidenschilling}}]{2005Icar..175..390T}
{Tegler}, S.~C., {Romanishin}, W., {Consolmagno}, G.~J., {et~al.} 2005,
  \icarus, 175, 390

\bibitem[{{Thirouin}(2013)}]{thirouin2013}
{Thirouin}, A. 2013, Ph.D. Thesis, University of Granada

\bibitem[{{Thirouin} {et~al.}(2010){Thirouin}, {Ortiz}, {Duffard},
  {Santos-Sanz}, {Aceituno}, \& {Morales}}]{2010A&A...522A..93T}
{Thirouin}, A., {Ortiz}, J.~L., {Duffard}, R., {et~al.} 2010, \aap, 522, A93

\bibitem[{{Thirouin} {et~al.}(2016){Thirouin}, {Sheppard}, {Noll}, {Moskovitz},
  {Ortiz}, \& {Doressoundiram}}]{2016AJ....151..148T}
{Thirouin}, A., {Sheppard}, S.~S., {Noll}, K.~S., {et~al.} 2016, \aj, 151, 148

\bibitem[{{Trujillo} {et~al.}(2007){Trujillo}, {Brown}, {Barkume}, {Schaller},
  \& {Rabinowitz}}]{2007ApJ...655.1172T}
{Trujillo}, C.~A., {Brown}, M.~E., {Barkume}, K.~M., {Schaller}, E.~L., \&
  {Rabinowitz}, D.~L. 2007, \apj, 655, 1172

\bibitem[{{Vilenius} {et~al.}(2012){Vilenius}, {Kiss}, {Mommert}, {M{\"u}ller},
  {Santos-Sanz}, {Pal}, {Stansberry}, {Mueller}, {Peixinho}, {Fornasier},
  {Lellouch}, {Delsanti}, {Thirouin}, {Ortiz}, {Duffard}, {Perna}, {Szalai},
  {Protopapa}, {Henry}, {Hestroffer}, {Rengel}, {Dotto}, \&
  {Hartogh}}]{2012A&A...541A..94V}
{Vilenius}, E., {Kiss}, C., {Mommert}, M., {et~al.} 2012, \aap, 541, A94

\bibitem[{{Vilenius} {et~al.}(2014){Vilenius}, {Kiss}, {M{\"u}ller}, {Mommert},
  {Santos-Sanz}, {P{\'a}l}, {Stansberry}, {Mueller}, {Peixinho}, {Lellouch},
  {Fornasier}, {Delsanti}, {Thirouin}, {Ortiz}, {Duffard}, {Perna}, \&
  {Henry}}]{2014A&A...564A..35V}
{Vilenius}, E., {Kiss}, C., {M{\"u}ller}, T., {et~al.} 2014, \aap, 564, A35

\end{thebibliography}

%______________________________________________________________   

\Online
%______________________________________________________________   

\begin{appendix} %First online appendix

\onecolumn

\section{Tables}
%\label{Datared}
\label{append1}
In this appendix, we include the light curve photometric results obtained from the Herschel/PACS observations of Haumea, 2003 VS$_{2}$ and 2003 AZ$_{84}$.

\begin{longtable}{cccc}

\caption{Haumea thermal time series photometry results from Herschel/PACS observations at green (100 $\mu$m) and red (160 $\mu$m) bands (some clear outliers have been removed in the table). Each data point in the green band spans around 18.8 minutes and is the combination of four single images, there is an overlap of around 14.1 minutes between consecutive data points with the same OBSID (except when some outliers have been removed). For the red band each data point spans around 28.2 minutes and is the combination of six single frames, there is an overlap of around 23.5 minutes between consecutive data points with the same OBSID (except when some outliers have been removed). Thermal light curves in Fig. \ref{LC_haumea_g_r} have been obtained from these data folding the dates with the Haumea's rotational period and computing a running mean with a temporal bin of 0.05 in rotational phase. OBSID are the Herschel internal observation IDs, JD are the julian dates at the middle of the integration uncorrected for light-time (the mean one-way light-time for OBSID 1342188470 is 426.329455 min, and 421.967915 min for OBSID 1342198851), Band are the different filters (green or red) used to observe with PACS, Flux/unc are the in band fluxes and 1-$\sigma$ associated uncertainties expresed in millijansky (mJy), these values must be divided by the factors 0.98 and 0.99 for the green and red bands, respectively, to obtain color corrected fluxes/uncertainties. The zero time used to fold the rotational light curves in Fig. \ref{LC_haumea_g_r} is JD = 2455188.720000 days (uncorrected for light-time, the one-way light-time for this date is 426.308537 min).}\\ % title of Table

\hline\hline % inserts double horizontal lines

OBSID   &       JD      & Band   &      Flux/unc \\
        & [days] &       &       [mJy]  \\

% table heading

\hline % inserts single horizontal line

\endfirsthead  

\multicolumn{4}{c}{\tablename\ \thetable\ -- Haumea thermal time series with Herschel/PACS \textit{Continued from previous page}} \\

\hline
\hline
OBSID   &       JD      & Band   &      Flux/unc \\
        & [days] &       &       [mJy]  \\
\hline
\endhead
\hline \multicolumn{4}{c}{\textit{Continued on next page}} \\
\endfoot
\hline
\endlastfoot

\label{HaumeaPACSdata} % is used to refer this table in the text

1342188470 & 2455188.7448774963 & green & 23.61 $\pm$ 2.38 \\
1342188470 & 2455188.7537522637 & green & 24.44 $\pm$ 2.36 \\
1342188470 & 2455188.7579938867 & green & 24.94 $\pm$ 2.39 \\
1342188470 & 2455188.7622028766 & green & 26.64 $\pm$ 2.33 \\
1342188470 & 2455188.7663792432 & green & 27.80 $\pm$ 2.51 \\
1342188470 & 2455188.7705882331 & green & 26.12 $\pm$ 2.66 \\
1342188470 & 2455188.7740794080 & green & 24.05 $\pm$ 2.46 \\
1342188470 & 2455188.7775705927 & green & 22.78 $\pm$ 2.43 \\
1342188470 & 2455188.7810617676 & green & 20.10 $\pm$ 2.41 \\
1342188470 & 2455188.7845529523 & green & 18.72 $\pm$ 2.29 \\
1342188470 & 2455188.7880441272 & green & 15.87 $\pm$ 2.29 \\
1342188470 & 2455188.7915353021 & green & 14.35 $\pm$ 2.37 \\
1342188470 & 2455188.7950591105 & green & 16.87 $\pm$ 2.26 \\
1342188470 & 2455188.7985502952 & green & 16.35 $\pm$ 2.38 \\
1342188470 & 2455188.8020414701 & green & 15.89 $\pm$ 2.41 \\
1342188470 & 2455188.8055326547 & green & 16.49 $\pm$ 2.43 \\
1342188470 & 2455188.8090238296 & green & 17.20 $\pm$ 2.42 \\
1342188470 & 2455188.8125150045 & green & 15.67 $\pm$ 2.71 \\
1342188470 & 2455188.8160061892 & green & 13.90 $\pm$ 2.60 \\
1342188470 & 2455188.8188121826 & green & 14.09 $\pm$ 2.64 \\
1342188470 & 2455188.8209003611 & green & 14.30 $\pm$ 2.65 \\
1342188470 & 2455188.8286331687 & green & 16.71 $\pm$ 2.22 \\
1342188470 & 2455188.8307213471 & green & 20.04 $\pm$ 2.18 \\
1342188470 & 2455188.8334947070 & green & 23.30 $\pm$ 2.19 \\
1342188470 & 2455188.8369858819 & green & 26.45 $\pm$ 2.24 \\
1342188470 & 2455188.8404770764 & green & 27.89 $\pm$ 2.19 \\
1342188470 & 2455188.8440008848 & green & 28.96 $\pm$ 2.07 \\
1342188470 & 2455188.8474920597 & green & 29.05 $\pm$ 2.17 \\
1342188470 & 2455188.8509832346 & green & 27.61 $\pm$ 2.21 \\
1342188470 & 2455188.8544744095 & green & 27.16 $\pm$ 2.29 \\
1342188470 & 2455188.8579655844 & green & 22.66 $\pm$ 2.52 \\
1342188470 & 2455188.8614567788 & green & 22.28 $\pm$ 2.76 \\
1342188470 & 2455188.8649479537 & green & 22.19 $\pm$ 2.84 \\
1342188470 & 2455188.8684717622 & green & 22.00 $\pm$ 2.86 \\
1342188470 & 2455188.8719629371 & green & 20.67 $\pm$ 2.79 \\
1342188470 & 2455188.8754541120 & green & 23.04 $\pm$ 2.83 \\
1342188470 & 2455188.8835458173 & green & 19.90 $\pm$ 2.82 \\

1342198851 & 2455368.3686949732 & green & 24.11 $\pm$ 2.01 \\
1342198851 & 2455368.3700889852 & green & 23.14 $\pm$ 2.00 \\
1342198851 & 2455368.3720406014 & green & 23.52 $\pm$ 1.95 \\
1342198851 & 2455368.3740121326 & green & 22.45 $\pm$ 1.93 \\
1342198851 & 2455368.3759239200 & green & 22.81 $\pm$ 2.02 \\
1342198851 & 2455368.3778954512 & green & 23.87 $\pm$ 2.00 \\
1342198851 & 2455368.3804644155 & green & 23.92 $\pm$ 1.98 \\
1342198851 & 2455368.3830532948 & green & 23.42 $\pm$ 2.03 \\
1342198851 & 2455368.3856620882 & green & 22.77 $\pm$ 2.01 \\
1342198851 & 2455368.3889081441 & green & 21.12 $\pm$ 2.03 \\
1342198851 & 2455368.3921741145 & green & 21.51 $\pm$ 2.12 \\
1342198851 & 2455368.3954400853 & green & 21.56 $\pm$ 2.25 \\
1342198851 & 2455368.3987060557 & green & 20.84 $\pm$ 2.26 \\
1342198851 & 2455368.4019720261 & green & 20.37 $\pm$ 2.43 \\
1342198851 & 2455368.4052379965 & green & 21.14 $\pm$ 2.31 \\
1342198851 & 2455368.4085039669 & green & 19.01 $\pm$ 2.49 \\
1342198851 & 2455368.4117500233 & green & 17.80 $\pm$ 2.28 \\
1342198851 & 2455368.4150159936 & green & 17.76 $\pm$ 2.18 \\
1342198851 & 2455368.4182819640 & green & 18.46 $\pm$ 2.15 \\
1342198851 & 2455368.4215479344 & green & 18.03 $\pm$ 2.29 \\
1342198851 & 2455368.4248139048 & green & 18.24 $\pm$ 2.26 \\
1342198851 & 2455368.4287171378 & green & 19.91 $\pm$ 2.44 \\
1342198851 & 2455368.4319831086 & green & 21.60 $\pm$ 2.66 \\
1342198851 & 2455368.4345919020 & green & 22.69 $\pm$ 2.65 \\
1342198851 & 2455368.4431551173 & green & 23.93 $\pm$ 2.05 \\
1342198851 & 2455368.4451266481 & green & 24.54 $\pm$ 1.89 \\
1342198851 & 2455368.4477752708 & green & 23.71 $\pm$ 1.91 \\
1342198851 & 2455368.4510810701 & green & 23.69 $\pm$ 1.92 \\
1342198851 & 2455368.4543271260 & green & 24.32 $\pm$ 2.00 \\
1342198851 & 2455368.4575930964 & green & 25.63 $\pm$ 2.24 \\
1342198851 & 2455368.4608590668 & green & 25.58 $\pm$ 2.46 \\
1342198851 & 2455368.4641250377 & green & 26.49 $\pm$ 2.53 \\
1342198851 & 2455368.4673910080 & green & 25.53 $\pm$ 2.28 \\
1342198851 & 2455368.4706569784 & green & 23.38 $\pm$ 2.25 \\
1342198851 & 2455368.4739229488 & green & 22.80 $\pm$ 2.17 \\
1342198851 & 2455368.4771690047 & green & 21.66 $\pm$ 2.07 \\
1342198851 & 2455368.4804349756 & green & 19.40 $\pm$ 1.89 \\
1342198851 & 2455368.4837009460 & green & 17.78 $\pm$ 2.03 \\
1342198851 & 2455368.4869669164 & green & 17.91 $\pm$ 2.08 \\
1342198851 & 2455368.4902328867 & green & 17.52 $\pm$ 2.12 \\
1342198851 & 2455368.4934988571 & green & 14.78 $\pm$ 2.27 \\
1342198851 & 2455368.4967449135 & green & 16.90 $\pm$ 2.43 \\
1342198851 & 2455368.5000108839 & green & 16.57 $\pm$ 2.37 \\
1342198851 & 2455368.5032768543 & green & 15.93 $\pm$ 2.25 \\
1342198851 & 2455368.5065428247 & green & 15.26 $\pm$ 2.28 \\
1342198851 & 2455368.5098087951 & green & 19.69 $\pm$ 2.13 \\
1342198851 & 2455368.5130747659 & green & 20.06 $\pm$ 2.23 \\
1342198851 & 2455368.5163208218 & green & 20.92 $\pm$ 2.24 \\
1342198851 & 2455368.5195867922 & green & 22.20 $\pm$ 2.28 \\
1342198851 & 2455368.5227531902 & green & 24.30 $\pm$ 2.21 \\
1342198851 & 2455368.5253619840 & green & 22.66 $\pm$ 2.37 \\
1342198851 & 2455368.5279508629 & green & 23.96 $\pm$ 2.27 \\
1342198851 & 2455368.5299024796 & green & 24.67 $\pm$ 2.14 \\

\hline  

1342188470 & 2455188.7545345956 & red & 23.15 $\pm$ 4.10 \\
1342188470 & 2455188.7583467197 & red & 24.92 $\pm$ 4.20 \\
1342188470 & 2455188.7618420459 & red & 22.16 $\pm$ 3.10 \\
1342188470 & 2455188.7653373731 & red & 24.36 $\pm$ 2.30 \\
1342188470 & 2455188.7688332782 & red & 25.14 $\pm$ 2.50 \\
1342188470 & 2455188.7723280261 & red & 24.99 $\pm$ 2.50 \\
1342188470 & 2455188.7758239312 & red & 22.66 $\pm$ 2.40 \\
1342188470 & 2455188.7793186796 & red & 19.05 $\pm$ 3.20 \\
1342188470 & 2455188.7828145847 & red & 17.91 $\pm$ 2.60 \\
1342188470 & 2455188.7863093326 & red & 18.62 $\pm$ 4.20 \\
1342188470 & 2455188.7898052381 & red & 16.14 $\pm$ 4.50 \\
1342188470 & 2455188.7933005644 & red & 18.69 $\pm$ 4.20 \\
1342188470 & 2455188.7967958916 & red & 18.69 $\pm$ 4.10 \\
1342188470 & 2455188.8002912179 & red & 20.25 $\pm$ 3.70 \\
1342188470 & 2455188.8037865451 & red & 18.13 $\pm$ 3.50 \\
1342188470 & 2455188.8072818713 & red & 22.59 $\pm$ 2.80 \\
1342188470 & 2455188.8107771981 & red & 16.71 $\pm$ 3.50 \\
1342188470 & 2455188.8142731031 & red & 17.21 $\pm$ 3.90 \\
1342188470 & 2455188.8177678520 & red & 20.53 $\pm$ 3.80 \\
1342188470 & 2455188.8212637566 & red & 20.75 $\pm$ 4.50 \\
1342188470 & 2455188.8247585054 & red & 21.38 $\pm$ 4.30 \\
1342188470 & 2455188.8282544101 & red & 21.67 $\pm$ 4.30 \\
1342188470 & 2455188.8317491589 & red & 20.32 $\pm$ 4.10 \\
1342188470 & 2455188.8352450635 & red & 19.97 $\pm$ 4.00 \\
1342188470 & 2455188.8387403907 & red & 22.59 $\pm$ 4.00 \\
1342188470 & 2455188.8422357170 & red & 20.82 $\pm$ 3.30 \\
1342188470 & 2455188.8457310442 & red & 22.37 $\pm$ 3.30 \\
1342188470 & 2455188.8492263705 & red & 24.07 $\pm$ 2.40 \\
1342188470 & 2455188.8527216977 & red & 23.93 $\pm$ 2.50 \\
1342188470 & 2455188.8562170244 & red & 25.70 $\pm$ 3.10 \\
1342188470 & 2455188.8597129295 & red & 22.59 $\pm$ 2.80 \\
1342188470 & 2455188.8632076778 & red & 24.92 $\pm$ 2.40 \\
1342188470 & 2455188.8667035829 & red & 24.07 $\pm$ 3.30 \\
1342188470 & 2455188.8701983313 & red & 22.73 $\pm$ 3.30 \\
1342188470 & 2455188.8736044089 & red & 21.31 $\pm$ 4.00 \\

1342198851 & 2455368.3739723968 & red & 25.55 $\pm$ 3.30 \\
1342198851 & 2455368.3775566947 & red & 24.42 $\pm$ 3.00 \\
1342198851 & 2455368.3808211582 & red & 24.10 $\pm$ 4.20 \\
1342198851 & 2455368.3840844650 & red & 24.42 $\pm$ 4.00 \\
1342198851 & 2455368.3873489285 & red & 23.37 $\pm$ 5.00 \\
1342198851 & 2455368.3906122353 & red & 23.29 $\pm$ 4.90 \\
1342198851 & 2455368.3938766997 & red & 23.37 $\pm$ 4.30 \\
1342198851 & 2455368.3971400065 & red & 23.86 $\pm$ 3.50 \\
1342198851 & 2455368.4004044710 & red & 23.78 $\pm$ 2.90 \\
1342198851 & 2455368.4036677782 & red & 24.42 $\pm$ 2.80 \\
1342198851 & 2455368.4069322422 & red & 25.07 $\pm$ 2.40 \\
1342198851 & 2455368.4101955500 & red & 24.99 $\pm$ 2.50 \\
1342198851 & 2455368.4134600144 & red & 23.13 $\pm$ 2.30 \\
1342198851 & 2455368.4167233212 & red & 20.94 $\pm$ 2.30 \\
1342198851 & 2455368.4199877856 & red & 19.00 $\pm$ 3.10 \\
1342198851 & 2455368.4232510934 & red & 14.31 $\pm$ 3.10 \\
1342198851 & 2455368.4265155573 & red & 12.37 $\pm$ 3.10 \\
1342198851 & 2455368.4297788651 & red &  9.22 $\pm$ 3.20 \\
1342198851 & 2455368.4330433300 & red & 11.48 $\pm$ 2.30 \\
1342198851 & 2455368.4363066372 & red & 11.97 $\pm$ 2.60 \\
1342198851 & 2455368.4395711031 & red & 14.88 $\pm$ 2.30 \\
1342198851 & 2455368.4428344108 & red & 17.63 $\pm$ 3.10 \\
1342198851 & 2455368.4460988762 & red & 16.82 $\pm$ 2.60 \\
1342198851 & 2455368.4493621849 & red & 19.08 $\pm$ 3.50 \\
1342198851 & 2455368.4526266507 & red & 18.92 $\pm$ 3.40 \\
1342198851 & 2455368.4558899594 & red & 21.03 $\pm$ 3.60 \\
1342198851 & 2455368.4591544257 & red & 20.94 $\pm$ 3.80 \\
1342198851 & 2455368.4624177348 & red & 20.22 $\pm$ 3.40 \\
1342198851 & 2455368.4656822011 & red & 21.11 $\pm$ 3.60 \\
1342198851 & 2455368.4689455107 & red & 22.56 $\pm$ 3.30 \\
1342198851 & 2455368.4722099770 & red & 23.37 $\pm$ 3.10 \\
1342198851 & 2455368.4754732866 & red & 20.78 $\pm$ 3.20 \\
1342198851 & 2455368.4787377538 & red & 19.81 $\pm$ 3.40 \\
1342198851 & 2455368.4820010634 & red & 16.42 $\pm$ 3.90 \\
1342198851 & 2455368.4852655306 & red & 17.63 $\pm$ 4.30 \\
1342198851 & 2455368.4885288402 & red & 17.55 $\pm$ 5.00 \\
1342198851 & 2455368.4917933075 & red & 17.47 $\pm$ 3.40 \\
1342198851 & 2455368.4950566171 & red & 18.68 $\pm$ 3.60 \\
1342198851 & 2455368.4983210848 & red & 19.49 $\pm$ 4.20 \\
1342198851 & 2455368.5015843948 & red & 17.95 $\pm$ 3.20 \\
1342198851 & 2455368.5048488625 & red & 15.45 $\pm$ 2.60 \\
1342198851 & 2455368.5081121726 & red & 18.20 $\pm$ 2.70 \\
1342198851 & 2455368.5113766398 & red & 16.25 $\pm$ 3.00 \\
1342198851 & 2455368.5146399504 & red & 16.82 $\pm$ 3.20 \\
1342198851 & 2455368.5179026825 & red & 17.14 $\pm$ 2.90 \\
1342198851 & 2455368.5211665714 & red & 19.81 $\pm$ 3.00 \\
1342198851 & 2455368.5244304603 & red & 19.41 $\pm$ 4.30 \\
1342198851 & 2455368.5276943492 & red & 18.20 $\pm$ 3.50 \\
1342198851 & 2455368.5309582381 & red & 17.87 $\pm$ 2.80 \\
1342198851 & 2455368.5341299847 & red & 15.36 $\pm$ 2.80 \\

\hline %inserts single line

%\footnotesize{}

\end{longtable}

%-------------------------------------------------------------------------------------------

\begin{longtable}{cccc}

\caption{2003 VS$_{2}$ thermal time series observations with Herschel/PACS at blue (70 $\mu$m) and red (160 $\mu$m) bands (some clear outliers have been removed in the table). Each data point in the blue band spans around 23.5 minutes and is the combination of five single images, there is no time overlap between consecutive images. For the red band each data point spans around 47 minutes and is the combination of ten single frames, there is an overlap of around 23.5 minutes between consecutive data points (except for the outliers removed). Thermal light curves in Fig. \ref{LC_vs2blue_red} are obtained folding these data with the 2003 VS$_{2}$ rotational period. OBSID is the Herschel internal observation ID, JD are the julian dates at the middle of the integration uncorrected for light-time (the mean one-way light-time is 306.228046 min), Band are the different filters (blue or red) used to observe with PACS, Flux/unc are the in band fluxes and 1-$\sigma$ associated uncertainties expressed in millijansky (mJy), these values must be divided by the factors 0.98 and 1.01 for the blue and red bands, respectively, to obtain color corrected fluxes/uncertainties. The zero time used to fold the rotational light curves in Fig. \ref{LC_vs2blue_red} is JD = 2452992.768380 days (uncorrected for light-time, the one-way light-time for this date is 296.244149 min).
}\\ % title of Table

\hline\hline % inserts double horizontal lines

OBSID   &       JD      & Band   &      Flux/unc \\
        & [days] &       &       [mJy]  \\

% table heading

\hline % inserts single horizontal line

\endfirsthead  

\multicolumn{4}{c}{\tablename\ \thetable\ -- 2003 VS$_{2}$ thermal time series with Herschel/PACS \textit{Continued from previous page}} \\

\hline
\hline
OBSID   &       JD      & Band   &      Flux/unc \\
        & [days] &       &       [mJy]  \\
\hline
\endhead
\hline \multicolumn{4}{c}{\textit{Continued on next page}} \\
\endfoot
\hline
\endlastfoot

\label{VS2PACSdata} % is used to refer this table in the text

1342202371 & 2455418.9107945664 & blue & 12.50 $\pm$ 1.80 \\ 
1342202371 & 2455418.9303779081 & blue & 13.60 $\pm$ 1.70 \\ 
1342202371 & 2455418.9466973604 & blue & 14.40 $\pm$ 1.70 \\ 
1342202371 & 2455418.9630168136 & blue & 15.90 $\pm$ 1.50 \\ 
1342202371 & 2455418.9793362664 & blue & 13.10 $\pm$ 1.90 \\ 
1342202371 & 2455418.9956557201 & blue & 13.20 $\pm$ 1.60 \\ 
1342202371 & 2455419.0152390650 & blue & 13.50 $\pm$ 2.00 \\ 
1342202371 & 2455419.0413501919 & blue & 14.60 $\pm$ 1.80 \\ 
1342202371 & 2455419.0576696461 & blue & 12.90 $\pm$ 1.60 \\ 
1342202371 & 2455419.0739890989 & blue & 15.30 $\pm$ 1.80 \\ 
1342202371 & 2455419.0903085498 & blue & 13.90 $\pm$ 1.70 \\ 
1342202371 & 2455419.1098918878 & blue & 15.80 $\pm$ 1.50 \\ 
1342202371 & 2455419.1262113363 & blue & 14.00 $\pm$ 1.70 \\ 
1342202371 & 2455419.1425307849 & blue & 14.00 $\pm$ 1.40 \\ 
1342202371 & 2455419.1588502331 & blue & 13.30 $\pm$ 1.50 \\ 
1342202371 & 2455419.1751696821 & blue & 12.70 $\pm$ 1.80 \\ 
1342202371 & 2455419.1914891317 & blue & 13.80 $\pm$ 1.20 \\ 
1342202371 & 2455419.2078085812 & blue & 14.20 $\pm$ 1.70 \\

\hline

1342202371 & 2455418.9254817832  & red &  8.30 $\pm$ 2.50 \\ 
1342202371 & 2455418.9418018144  & red &  9.60 $\pm$ 1.80 \\ 
1342202371 & 2455418.9581206883  & red & 13.70 $\pm$ 1.90 \\ 
1342202371 & 2455418.9744407199  & red & 13.30 $\pm$ 1.80 \\ 
1342202371 & 2455418.9907595953  & red & 10.80 $\pm$ 3.10 \\  
1342202371 & 2455419.0070796274  & red &  9.80 $\pm$ 1.40 \\ 
1342202371 & 2455419.0201351903  & red &  9.00 $\pm$ 2.20 \\ 
1342202371 & 2455419.0364540662  & red & 11.10 $\pm$ 3.10 \\ 
1342202371 & 2455419.0527740987  & red & 10.20 $\pm$ 2.90 \\ 
1342202371 & 2455419.0690929731  & red & 11.70 $\pm$ 2.40 \\ 
1342202371 & 2455419.0854130024  & red & 12.00 $\pm$ 1.60 \\ 
1342202371 & 2455419.1017318745  & red & 14.50 $\pm$ 1.90 \\ 
1342202371 & 2455419.1180519015  & red & 11.50 $\pm$ 2.40 \\ 
1342202371 & 2455419.1343707712  & red & 10.00 $\pm$ 1.40 \\ 
1342202371 & 2455419.1506907986  & red & 12.00 $\pm$ 2.00 \\ 
1342202371 & 2455419.1670096684  & red & 12.20 $\pm$ 2.30 \\ 
1342202371 & 2455419.1833296963  & red & 12.60 $\pm$ 1.60 \\ 
1342202371 & 2455419.1996485675  & red & 14.70 $\pm$ 2.50 \\ 

\hline %inserts single line

%\footnotesize{}

\end{longtable}

%-------------------------------------------------------------------------------------------

\begin{longtable}{cccc}

\caption{2003 AZ$_{84}$ thermal time series observations with Herschel/PACS at green (100 $\mu$m) band. Each data point spans around 28.2 minutes and is the combination of six single images, there is no time overlap between consecutive data points. Thermal light curve in Fig. \ref{LC_az84green} is obtained folding these data with the 2003 AZ$_{84}$ rotational period. OBSID are the Herschel internal observation IDs, JD are the julian dates at the middle of the integration uncorrected for light-time (the mean one-way light-time is 379.855579 min), Band is the filter used to observe with PACS, Flux/unc are the in band fluxes and 1-$\sigma$ associated uncertainties expressed in millijansky (mJy), these values must be divided by the factor 0.98 to obtain color corrected fluxes/uncertainties. The zero time used to fold the light curves in Fig. \ref{LC_az84green} is JD = 2453026.546400 days (uncorrected for light-time, the one-way light-time for this date is 373.264731 min).}\\ % title of Table

\hline\hline % inserts double horizontal lines

OBSID   &       JD      & Band   &      Flux/unc \\
        & [days] &       &       [mJy]  \\

% table heading

\hline % inserts single horizontal line

\endfirsthead  

\multicolumn{4}{c}{\tablename\ \thetable\ -- 2003 AZ$_{84}$ thermal time series with Herschel/PACS \textit{Continued from previous page}} \\

\hline
\hline
OBSID   &       JD      & Band   &      Flux/unc \\
        & [days] &       &       [mJy]  \\
\hline
\endhead
\hline \multicolumn{4}{c}{\textit{Continued on next page}} \\
\endfoot
\hline
\endlastfoot

\label{AZ84PACSdata} % is used to refer this table in the text

1342205152 & 2455466.5342869759 & green & 24.95 $\pm$ 2.17\\ 
1342205152 & 2455466.5538702821 & green & 27.07 $\pm$ 2.26\\ 
1342205152 & 2455466.5734535865 & green & 26.93 $\pm$ 1.98\\ 
1342205152 & 2455466.5930368891 & green & 28.81 $\pm$ 2.05\\ 
1342205152 & 2455466.6126201935 & green & 24.93 $\pm$ 2.40\\ 
1342205152 & 2455466.6224124241 & green & 27.77 $\pm$ 2.49\\ 
1342205152 & 2455466.6419957289 & green & 26.28 $\pm$ 2.43\\ 
1342205152 & 2455466.6615790343 & green & 29.85 $\pm$ 3.05\\ 
1342205152 & 2455466.6811623396 & green & 24.50 $\pm$ 2.76\\ 
1342205152 & 2455466.7007456459 & green & 26.36 $\pm$ 2.33\\ 
1342205152 & 2455466.7203289527 & green & 31.56 $\pm$ 1.71\\ 
1342205152 & 2455466.7399122599 & green & 27.40 $\pm$ 2.75\\ 
1342205152 & 2455466.7594955675 & green & 27.24 $\pm$ 3.43\\ 
1342205152 & 2455466.7790788747 & green & 26.24 $\pm$ 2.44\\ 
1342205152 & 2455466.7969396170 & green & 27.96 $\pm$ 2.76\\ 

\hline %inserts single line

%\footnotesize{}

\end{longtable}

\twocolumn

\end{appendix}

\end{document}